\documentclass[reprint,floatfix,notitlepage,nofootinbib,twocolumn,superscriptaddress,pra,preprintnumbers]{revtex4-1}
\usepackage{physics}
\usepackage{float}
\usepackage{amsmath}
\usepackage{amssymb}
\usepackage{graphicx}
\usepackage{enumitem}
\usepackage{pifont,bbm}
\usepackage[tight]{subfigure}
\usepackage{xifthen}
\usepackage{xargs}
\usepackage{tabularray}
\usepackage{multirow}
\usepackage{makecell}
\usepackage{xspace}
\usepackage{tikz}
\usetikzlibrary{calc,fadings,decorations.pathreplacing,shapes,shapes.multipart,arrows,shapes.misc,intersections,positioning}
\usepackage{xifthen}
\usepackage{xargs}

\newcommandx{\fineq}[4][1=-.8ex,2=1,3=1]{
  \begin{tikzpicture}[baseline={([yshift=#1]current  bounding  box.center)}, scale = #2, every node/.style={scale = #3}]
    #4
  \end{tikzpicture}
}

\newcommandx{\lrbd}[1][1=1]{
  \fineq[-0.8ex][0.08][0.2]{
    \fill (0,1)--(0,4)--(1,4)--cycle;
    \fill (2,2)--(2,4)--(1,4)--cycle;
  }
}
\newcommandx{\bbd}[1][1=1]{
  \fineq[-0.8ex][0.06][0.2]{
    \fill (0,1)--(1,4)--(2,1.5)--cycle;
  }
}

\newcommandx{\drawbox}[6][1=0,2=0,3=1,4=1,5=,6=]{
  \ifthenelse{\equal{#5}{}}{
    \draw[line width = 0.5pt] (#1,#2) rectangle (#3,#4);
  }{
    \draw[line width = 0.5pt, fill = #5] (#1,#2) rectangle (#3,#4);
  }
  \node () at (#1*0.5+#3*0.5,#2*0.5+#4*0.5) {#6};
}

\newcommandx{\regbox}[4][1=0,2=0,3=,4=]{
  \begin{scope}[shift={(#1,#2)}]
    \drawbox[0][0][1.5][1][#3][#4];
  \end{scope}
}

\newcommandx{\sqzbox}[5][1=0,2=0,3=,4=,5=]{
  \begin{scope}[shift={(#1,#2)}]
    \drawbox[0][0][2.25][1][#3][#4];
    \ifthenelse{\equal{#5}{s}}{
      \draw (0,0.1)--++(-0.1,0)--++(0,1)--++(2.25,0)--++(0,-0.1);
    }{}
  \end{scope}
}

\newcommandx{\sufourpart}[5][1=0,2=0,3=0.25,4=0.5,5=]{
  \begin{scope}[shift={(#1,#2)}]
    \pgfmathsetmacro{\r}{#4};
    \pgfmathsetmacro{\cent}{#3+#4};
    \draw (0,0)--++(0,#3);
    \draw (-\r,\cent-\r) rectangle (\r,\cent+\r);
    \draw (0,\cent+\r)--++(0,#3);
    \ifthenelse{\equal{#5}{}}{}{
      \node () at (0,\cent) {#5};
    }
  \end{scope}
}

\newcommandx{\sufour}[5][1=,2=,3=,4=,5=]
{
  \fineq[-0.8ex][0.6][1]{
    \sqzbox[0][0][][$#1$];
    \sufourpart[0.5][1][0.25][0.5][$#2$];
    \sufourpart[1.75][1][0.25][0.5][$#3$];
    \sufourpart[0.5][-1.5][0.25][0.5][$#4$];
    \sufourpart[1.75][-1.5][0.25][0.5][$#5$];
  }
}

\newcommandx{\dualgate}[7][1=0,2=0,3=,4=,5=,6=,7=]{
  \begin{scope}[shift={(#1,#2)}]
    \ifthenelse{\equal{#3}{l} \OR \equal{#3}{}}{
      \draw (0,-0.2)--(0.2,-0.2)--(0.2,0.2)--(0,0.2);
      \draw (0.2,0.2)--(0.5,0.5);
      \node () at (0.6,0.6) {$#5$};
      \draw (0.2,-0.2)--(0.5,-0.5);
      \node () at (0.6,-0.6) {$#7$};
    }{}
    \ifthenelse{\equal{#3}{r} \OR \equal{#3}{}}{
      \draw (0,-0.2)--(-0.2,-0.2)--(-0.2,0.2)--(0,0.2);
      \draw (-0.2,0.2)--(-0.5,0.5);
      \node () at (-0.6,0.6) {$#4$};
      \draw (-0.2,-0.2)--(-0.5,-0.5);
      \node () at (-0.6,-0.6) {$#6$};
    }{}
  \end{scope}
}
\newcommandx{\dualgatefat}[7][1=0,2=0,3=,4=,5=,6=,7=]{
  \begin{scope}[shift={(#1,#2)}]
    \ifthenelse{\equal{#3}{l} \OR \equal{#3}{}}{
      \draw[line width=1pt] (0,-0.2)--(0.2,-0.2)--(0.2,0.2)--(0,0.2);
      \draw[line width=1pt] (0.2,0.2)--(0.5,0.5);
      \node () at (0.6,0.6) {$#5$};
      \draw[line width=1pt] (0.2,-0.2)--(0.5,-0.5);
      \node () at (0.6,-0.6) {$#7$};
    }{}
    \ifthenelse{\equal{#3}{r} \OR \equal{#3}{}}{
      \draw[line width=1pt] (0,-0.2)--(-0.2,-0.2)--(-0.2,0.2)--(0,0.2);
      \draw[line width=1pt] (-0.2,0.2)--(-0.5,0.5);
      \node () at (-0.6,0.6) {$#4$};
      \draw[line width=1pt] (-0.2,-0.2)--(-0.5,-0.5);
      \node () at (-0.6,-0.6) {$#6$};
    }{}
  \end{scope}
}

\usepackage{simpler-wick}

\usepackage[english]{babel}
\makeatletter
\def\bbl@set@language#1{%
  \edef\languagename{%
    \ifnum\escapechar=\expandafter`\string#1\@empty
    \else\string#1\@empty\fi}%
  \@ifundefined{babel@language@alias@\languagename}{}{%
    \edef\languagename{\@nameuse{babel@language@alias@\languagename}}%
  }%
  \select@language{\languagename}%
  \expandafter\ifx\csname date\languagename\endcsname\relax\else
    \if@filesw
      \protected@write\@auxout{}{\string\select@language{\languagename}}%
      \bbl@for\bbl@tempa\BabelContentsFiles{%
        \addtocontents{\bbl@tempa}{\xstring\select@language{\languagename}}}%
      \bbl@usehooks{write}{}%
    \fi
  \fi}
\newcommand{\DeclareLanguageAlias}[2]{%
  \global\@namedef{babel@language@alias@#1}{#2}%
}
\makeatother
\DeclareLanguageAlias{en}{english}

\newcommand{\bw}{brickwall\xspace}
\newcommand{\stair}{staircase\xspace}

\usepackage{bm}
\renewcommand{\vec}[1]{\boldsymbol{\mathbf{#1}}}

\newcommand{\appref}[1]{App.~\ref{#1}}

\usepackage[colorlinks=true,citecolor=blue,linkcolor=magenta]{hyperref}

\graphicspath{{./}{./images/}}
\preprint{MIT-CTP/5357}

\begin{document}
\title{A Physical Theory of Two-stage Thermalization}

\author{Cheryne Jonay}
\affiliation{Department of Physics, Stanford University, Stanford, CA 94305}

\author{Tianci Zhou}
\affiliation{Center for Theoretical Physics, Massachusetts Institute of Technology, Cambridge, Massachusetts 02139, USA}
\affiliation{Department of Physics, Virginia Tech, Blacksburg, Virginia 24061, USA}
 
\date{\today}

\begin{abstract}
One indication of thermalization time is subsystem entanglement reaching thermal values. Recent studies on local quantum circuits reveal two exponential stages with decay rates $r_1$ and $r_2$ of the purity before and after thermalization. We provide an entanglement membrane theory interpretation, with $r_1$ corresponding to the domain wall free energy. Circuit geometry can lead to $r_1 < r_2$, producing a ``phantom eigenvalue". Competition between the domain wall and magnon leads to $r_2 < r_1$ when the magnon prevails. 
However, when the domain wall wins, this mechanism provides a practical approach for measuring entanglement growth through local correlation functions.
\end{abstract}
\maketitle

\begin{figure}[h]
\centering
\includegraphics[width=0.53\columnwidth]{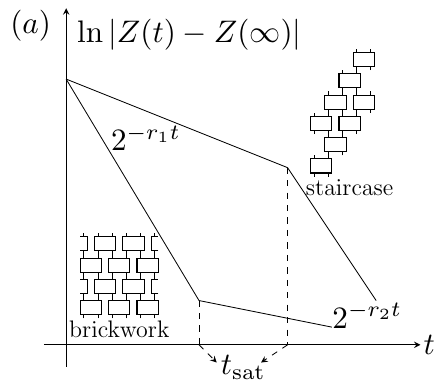}
\includegraphics[width=0.45\columnwidth]{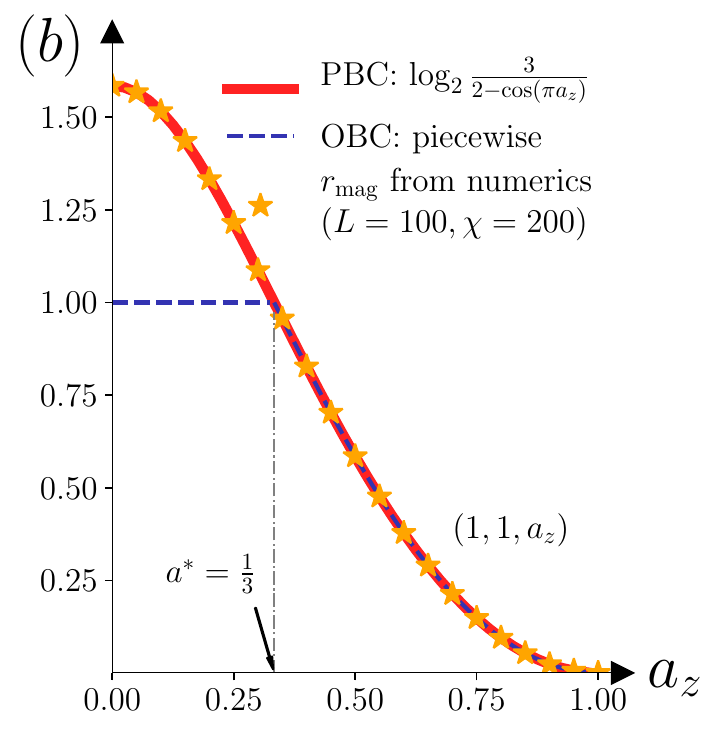}
\caption{Two-stage thermalization. (a): Schematics of 2-stage decay for \stair and \bw geometry. Both insets show circuits of 4 time steps. Generally $r_1 < r_2$ for \stair geometry and $r_1 \ge r_2$ for brickwork geometry. (b) For the dual unitary $(1,1,a_z)$ circuits, $r_2$ depends on $a_z$ and the boundary condition (periodic or open).  
}
\label{fig:schematics_r2az}
\end{figure}

Thermalization occurs in non-equilibrium states when local observables relax to their thermal equilibrium value. In a closed quantum system, this relaxation is achieved as the system locally approaches the maximal entropy states subject to the symmetry constraints\cite{srednicki_chaos_1994,deutsch_eigenstate_2018}. The entropy here is interpreted as entanglement entropy between subsystems. Its growth determines a characteristic time scale called “thermalization time”, after which the entanglement saturates. While the state at large may have equilibrated, dynamics at the microscopic scale continue. 
They only cause exponentially small fluctuations in the entanglement, but can induce drastic changes in other physical quantities. For example, the (computational) complexity of a quantum state continues to grow way beyond thermalization time (to quote 
“Entanglement is not enough”\cite{susskind_entanglement_2014}). Deviations from a fully thermal state can also be defined at a fixed time by looking at higher moments of the state: While the first moment may look thermal, higher moments do not. This coins the notion of “deep thermalization”\cite{ho_exact_2022,wilming_high-temperature_2022-1,lucas_generalized_2022,Cotler_emergent_state_design_2023,Choi_preparing_random_states_2023}, which seeks to define not only the distribution of states to approach thermal equilibrium, but also their higher moments. This is known as the state \textit{design} property\cite{DiVincenzo_quantum_data_hiding_2002,gross_evenly_2007,ambainis_quantum_t_designs_2007,Gross_structure_unitary_designs_2007,brandao_efficient_2016,Roberts_chaos_complexity_design_2017,Ippoliti_dynamical_purification_2023} in quantum information.\\ 
\indent This work addresses behaviors beyond the thermalization time in an  entanglement related quantity -- the state purity, defined as the trace of the reduced density matrix squared. It was recently found\cite{bensa_fastest_2021,znidaric_solvable_2022,bensa_purity_2022,bensa_two-step_2022,znidaric_phantom_2023,znidaric2023FloquetTwostep} that it generically hosts a multi-stage decay, the second stage taking place \textit{after} the system thermalizes. In a chaotic evolution, an initially pure subsystem increasingly loses resemblance to a pure state, which is evident in the decrease of the subsystem purity. For locally interacting systems, this decrease is exponential until the entanglement saturates (for a finite system) at a saturation time $t_{\rm sat}$\cite{Calabrese_evolution_ee_1dsystems_2005,Kim_ballistic_ee_2013,Nahum_2018,liu_entanglement_2014-1,casini_spread_2015}. Probing the two-stage thermalization therefore requires looking not at the decay of the purity itself, but the convergence towards its static saturation value at infinite time. 
To hint at its statistical mechanical nature, we denote the purity as a partition function $Z(t)$ and the quantity showing two stages $\ln ( | Z(t) - Z(\infty)| )$ (Fig.~\ref{fig:schematics_r2az}(a)). 
For qubits, the exponential decay is $2^{-r_1 t}$ before and $2^{-r_2 t}$ after saturation time.

In recent works, Refs.~\cite{bensa_fastest_2021,znidaric_solvable_2022} find that one can have $r_1 < r_2$ in circuits where the gates are stacked in a staircase geometry. This is a rather surprising effect if one use transfer matrix to compute $Z(t)$: it turns out the decay rate is not set by the second largest eigenvalue (the largest one is 1, which gives $Z(\infty)$ that has been subtracted off) , but instead by a ``phantom eigenvalue'' $2^{-r_2}$ that lies within the gap\cite{znidaric_phantom_2023}. To explain this phenomenon, Refs.~\cite{bensa_fastest_2021,znidaric_solvable_2022} appeal to the finite size of the subsystem and interpret it as a boundary effect in the framework of the transfer matrix. On the other hand, $r_1$ can also be greater than $r_2$ in a regular brickwork geometry (see Fig.~\ref{fig:schematics_r2az}). Although the ``phantom eigenvalue'' is absent, the mechanism preventing the ``mode'' of $r_2$ from appearing remains mysterious. 

In this work, we give an entanglement membrane\cite{zhou_entanglement_2020,jonay2018coarsegrained,mezei_entanglement_2017,mezei_exploring_2020,mezei_membrane_2018,gong_coarse-grained_2022} interpretation of the two-stage thermalization. When viewing the purity $Z(t)$ as a partition function of a emergent magnet, the two decay rates are the free energies of single particle modes. $r_1$ is associated with the free energy of the domain wall\cite{zhou_entanglement_2020,nahum_quantum_2017}. The ``phantom'' $r_1$ is created by a particular domain wall minimal path that exits in the staircase but not in the brickwall geometry; its appearance is a geometric effect. In both geometries, we find there is a magnon degree of freedom which can compete with the domain wall after $t> t_{\rm sat}$. If it wins, 
the free energy is lowered and $r_2 < r_1$. Our theory passes analytic checks for circuits with disorder averaging and the predictions are also confirmed in clean Floquet circuits with time translation symmetry. Based on this theory, we propose a novel experiment to measure entanglement from local correlation functions.

{\it Phenomenologies:} We first review the quantitative phenomenlogies observed for quantum circuits chosen by Ref.~\cite{bensa_fastest_2021}. The unitary gates in the circuits are nearest neighbour and stacked either in a \bw or   \stair geometry (Fig.~\ref{fig:schematics_r2az}(a)) on a one dimensional lattice of $L$ qudits with local Hilbert space dimension $q$. 
The \bw structure naturally arises in the Trotter limit and for modelling local interactions, which ensues a finite speed of information propagation.
The \stair geometry has the advantage that dynamics can be traced backwards along a single space-time path, which also reduces spatial cost in quantum simulation if reset is possible\cite{anikeeva_recycling_2021}. 
We identify three specific scenarios $r_1=r_2$, $r_1<r_2$ and $r_1>r_2$. 

{\it Scenario 0: $r_1=r_2$} We first consider the paradigmatic example of random unitary circuits (RUC) in which the gates are independent random ${\rm U}(q^2)$ matrices. This choice allows us to access the typical behavior via disorder averaging(see e.g. \cite{nahum2018operator,vonKeyserlingk2018operator,znidaric_exact_2008,harrow_random_2009,emerson_pseudo-random_2003,skinner2019measurement,nahum_quantum_2017,li2019measurement,chan2018spectral,chan_unitary-projective_2019} and a recent review\cite{fisher_random_2022}). We thus examine $\ln |\overline{Z(t)} - \overline{Z}(\infty)|$ where the over-line denotes averaging over Haar ensemble on ${\rm U}(q^2)$. In a \bw geometry, there is a single decay rate given by 
\begin{equation}
     r_1 = r_2 = \frac{\ln \frac{q^2 + 1}{2q} }{\ln q}, \hspace{.5cm} \text{[\bw, RUC].}
 \end{equation}

 
{\it Scenario 1: $r_1 < r_2$}. For RUC in a   \stair geometry, we obtain two decay rates after random averaging 
\cite{znidaric_solvable_2022}
\begin{equation}
\label{eq:r_1_r_2_S}
  r_1 = \frac{1}{2} \frac{\ln \frac{q^2 - q +1}{q}}{\ln q}, \quad r_2 = \frac{\ln \frac{q^2 + 1}{2q} }{\ln q}. \hspace{.5cm} \text{[S, RUC]}
\end{equation}
This phenomena is observed in more general (non-random)  circuits with \stair geometry. 


{\it Scenario 2: $r_1 > r_2$} \bw geometry can host different behaviors than scenario 0 if we allow more general gates than Haar random. Let us specialize to $q = 2$. An arbitrary 2-qubit gate can be parametrized through four single-qubit rotations $u_i,\> i=1,2,3,4$ 
and a symmetric 2-body interaction $u_{\rm sym}$ as  
\begin{equation}
u = (u_1 \otimes u_2 ) u_{\rm sym}  ( u_3 \otimes u_4). 
\end{equation}

The symmetric part is generated by commutative operators $u_{\rm sym}  = \exp( - i \frac{\pi}{4} (\sum_{\alpha = x, y, z} a_\alpha \sigma^\alpha \sigma^\alpha ) )$. Here $\sigma^{\alpha}$ is the $\alpha$-th Pauli matrix and $0 \le a_\alpha \le 1$. In this scenario, we average the four single-qubit unitaries $u_i$ over ${\rm U}(2)$, hence the whole gate is indexed by $(a_x, a_y, a_z)$. The choice $a_x = a_y = 1$ corresponds to a special class known as {\it dual unitary} gates, which means the gate is unitary viewed in both spatial and temporal directions (see e.g. \cite{bertini_exact_2019-1,bertini_entanglement_2018} and more details in a review\cite{prosen_many_2021}). We primarily focus on this one-parameter class of circuits, which we abbreviate as $(1,1,a_z)$. In the one parameter family $(1,1,a_z)$, $a_z = 0$ corresponds to a iSWAP gate; $a_z = 1$ a SWAP gate; the gate is not integrable elsewhere.
It has been proven that the purity decays as $\text{poly}(t) 2^{-t}$\cite{foligno_growth_2023} on average for dual unitary circuits with high enough entangling power, while numerics suggest it holds more generally. Consequently, the first stage decay rate is $r_1 = 1$ for OBC and $r_1=2$ for PBC (the subregion $A$ has two boundary points). The second stage decay rate $r_2$ is a function of $a_z$, 
\begin{equation}
\label{eq:r2_az_analy}
r_1 = 2, \quad r_2 (a_z) =  \frac{\ln \frac{3}{2 - \cos (\pi a_z)}}{\ln 2}, \hspace{.25cm} \text{[$(1,1,a_z)$, PBC]}. 
\end{equation}
We prove the expression of $r_2(a_z)$ in \eqref{eq:r2_az_analy} in \appref{app:AveragedChannel}, which was numerically computed and conjectured in \cite{bensa_two-step_2022} 
For OBC, the expression remains the same for $a_z \ge \frac{1}{3}$ but caps to $1$ for $0 \le a_z \le \frac{1}{3}$. See schematics in Fig.~\ref{fig:schematics_r2az}(b). The phenomenon of $r_1 > r_2$ extends beyond the $(a_x, a_y, a_z)$ circuits and applies even to Floquet circuits with space and time translation symmetries. 


{\it Effective magnet.} The physical theory originates from analyzing random unitary circuits, where averaged entanglement dynamics map to the statistical mechanics of an effective magnet, which we review now\cite{chan_soloution_2018,nahum_quantum_2017,zhou_nahum_emergent_stat_mech2018,zhou_entanglement_2020,Khemani_operator_hydro_2018,vonKeyserlingk2018operator,Vasseur_holographicTN_2019,Jian_mi_criticality_2020,Bao_theory_mipt_2020,hunterjones_unitary_designs_from_stat_mech_2019,Liu_ee_gravity_2021,Fisher_Vijay_CircuitsReview}. 
The relevant partition function for the problem at hand is given by the purity of a time evolved state restricted to subregion $A$,
\begin{align}
\label{eq:Z_A(t)}
    Z_A(t)= \Tr(\rho_A(t)^2), 
\end{align}
where $\rho_A(t) = \Tr_{\overline{A}}(\ket{\psi(t)} \bra{\psi(t)})$. This quantity contains two copies of the unitary circuit $U$ and its conjugate $U^{*}$, combines as $U\otimes U^{*} \otimes U \otimes U^{*}$ acting on four copies the $L$ qudits. Since our sampling of random gates is independent in space and time, the overall random averaging of $U$ reduces to the separate averaging over each individual gate. The local Haar average over the single-site unitary $\overline{u_1 \otimes u^*_1 \otimes u_1 \otimes u^*_1}$ projects the 4 copies of the qudit Hilbert space into a two-dimensional subspace spanned by the states $\{ |+ \rangle , |- \rangle \}$\cite{collins_integration_2006}, which represent our effective spins. 
Physically these two states denote two different ways to pair a unitary and its conjugate, 
\begin{align}
\label{eq:pairing}
\ket{+}&:= \,\, \wick{
    \c u_1 \otimes  \c u_1^* \otimes  \c u_1 \otimes  \c u_1^*  
  }, \\
\ket{-}&:= \,\, \wick{
    \c2 u_1 \otimes  \c1 u_1^* \otimes  \c1 u_1 \otimes  \c2 u_1^* 
  }
\end{align}
The boundary spins are specified by how different copies of $U$ and $U^*$ are connected together in the partition function Eq.~\ref{eq:Z_A(t)}: For half-system purity, the top boundary is a domain wall state $| +  \cdots + - \cdots - \rangle$ where region $\overline{A}$ ($A$) hosts $\ket{+}$  ($\ket{-}$) states; for an initial product state, the bottom boundary is free. 

\begin{figure}[h]
\centering
\includegraphics[width=0.95\columnwidth]{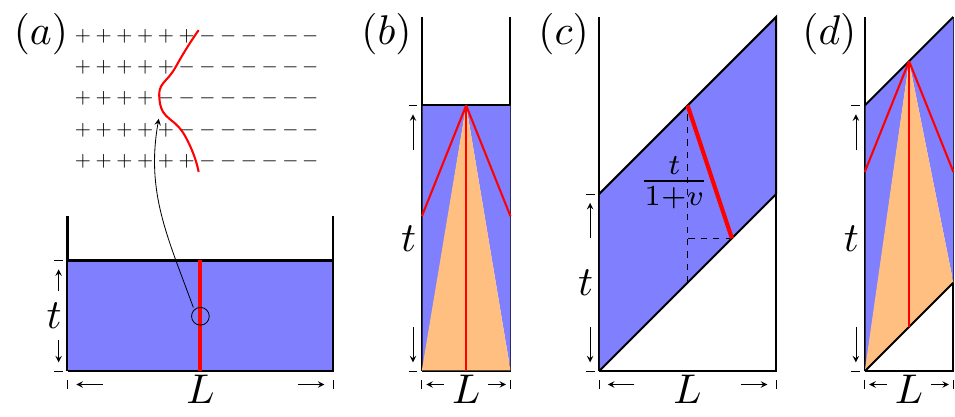}
\caption{Domain wall configurations in \bw ((a)(b)) and \stair ((c)(d)) geometries. (a) $t< t_{\rm sat}$, spin states at the lattice level (top) and domain wall random walk at the coarsed grained level (bottom). (b) $t> t_{\rm sat}$ for \bw. (c) $t< t_{\rm sat}$, domain wall has shorter paths in the \stair. (d) Same as (b) for \stair. }
\label{fig:dw_S}
\end{figure}

The local average $\overline{u \otimes u^* \otimes u \otimes u^*}$ is the local transfer matrix $M$ that determines the update rules of adjacent spins. For RUCs, the update rules are
\begin{align}
\ket{++} &\rightarrow \ket{++}  \label{eq:unitarity_constriant}\\
\ket{+-} &\rightarrow K \ket{++} + K \ket{--} \label{eq:K_move},
\end{align}
and symmetric counterparts by exchanging $\ket{+}$ and $\ket{-}$ with $K = \frac{2q}{q^2 + 1}$. Eq.~\eqref{eq:unitarity_constriant} arises from the unitarity of the gates, while Eq.~\eqref{eq:K_move} is the microscopic theory of a propagating domain walk. At each time step, the domain wall can move to the left or right with equal probability, but the global domain wall mode is preserved (Fig.~\ref{fig:dw_S}(a)) 
The free energy of this random walk is the R\'enyi entropy. The average partition function is exact $\overline{Z(t)} = (2K)^t$ \cite{zhou_nahum_emergent_stat_mech2018}, and the decay rate coincides with $r_2$ in Eq.~\eqref{eq:r_1_r_2_S}. For dual unitary circuits, the unitarity rules are the same as in Eq.~\eqref{eq:unitarity_constriant}. However, the domain wall \eqref{eq:K_move} is not preserved, and can be transferred into more states
\begin{equation}
\ket{+-} \rightarrow h \ket{++} + h\ket{--} + b_{+} \ket{+-} + b_{-} \ket{-+} \label{eq:RDUCmove}
\end{equation}
where $h=(3-v)/9$, $b_{\pm} = (3 \pm 6u + 5v)/36$ and $u = \cos ( \pi a_x) $$+ \cos ( \pi a_y ) + \cos ( \pi a_z) $, $v = \cos( \pi a_x) \cos( \pi a_y ) + \cos( \pi a_y) \cos( \pi a_z ) + \cos( \pi a_z) \cos( \pi a_x )$ \cite{bensa_fastest_2021}. For later reference, we denote the overall propagator $\mathcal{\widehat{M}}(t)$, which is an alternate product of the global transfer matrices at even and odd steps (\appref{app:Numerics}). The partition function is 
\begin{equation}
  Z(t) = \frac{1}{(q^2 + q)^L } \sum_{s_i = \pm } \langle s_1, \cdots s_L | \mathcal{\widehat{M}}(t)|   \cdots + - \cdots \rangle \label{eq:Z}.
\end{equation}
At scales much larger than the domain wall width, the membrane theory asserts that the free energy of the domain wall can be described by a macroscopic line tension function $\mathcal{E}(v)$, which only depends on the space-time (anti-)slope $v$. The partition function Eq.~\eqref{eq:Z} asymptotically decays as $\exp( - \min_v \mathcal{E}(v) t  \ln q )$. For RUCs, the average partition function has an exact line tension for a random walk: 
\begin{equation}
\label{eq:ruc_tension}
  \mathcal{E}_{}(v) = \frac{\ln \frac{q^2 + 1}{q} + \frac{1+v}{2} \ln \frac{1 + v}{2} + \frac{1- v}{2} \ln \frac{1 - v}{2} }{\ln q }. 
\end{equation}
The minimum at $v = 0$ is a vertical random walk (Fig.~\ref{fig:dw_S}(a)) and consistently gives $r_2$ in Eq.~\eqref{eq:r_1_r_2_S} of scenario 0.

{\it Scenario 1: $r_1 < r_2$ S-geometry}. The staircase geometry creates a tilted diamond region which constrains the domain wall's movement. As shown in Fig.~\ref{fig:dw_S}, the bottom staircase boundary is tilted upward at $45^\circ$, allowing the domain wall to travel a shorter distance if it tilts towards this lower boundary. However, this comes at the price of a larger line tension $\varepsilon(v)$. Quantitatively, if the domain wall has an (anti-)slope $v$, the distance traveled (or time duration) is $t/(1+v)$. The free energy $F(v)$ is a trade-off between shorter path length and increased line tension.
\begin{equation}
F(v) = \min_v \frac{\mathcal{E}(v)}{1+ v}\ln q. 
\end{equation}

For RUCs, the line tension is given in Eq.~\eqref{eq:ruc_tension}, and the minimum at $v^* = \frac{(q-1)^2}{q^2 + 1}$ reproduces $r_1 = \frac{\mathcal{E}(v^*)}{1+ v*} = \frac{1}{2} \frac{\ln \frac{q^2 - q + 1}{q}}{\ln q}$, which is consistent with Eq.~\eqref{eq:r_1_r_2_S}. However, for $t \sim t_{\rm sat}$, the same trajectory exits at the spatial boundary (rather than the bottom boundary). For $t > t_{\rm sat}$, the partition function has almost saturated to a static value. Schematically (Fig.~\ref{fig:dw_S}(d)), the saturation value comes from the contribution in which the domain wall hits the left and right boundaries. And the second stage decay come from a subleading correction in which the domain wall continues to reach the bottom (yellow). The effect of the tilted bottom boundary is increasingly negligible compared to the bulk contribution. Consequently, the yellow part is dominated by an (almost) vertical domain wall $e^{-\mathcal{E}(0) t \ln q }$, which gives the decay rate $r_2 = \mathcal{E}(0)$. Taking the tilt angle to zero reduces to the \bw geometry, which explains why the \bw RUCs exhibit no geometric cross-over (scenario 0).  


\textit{Scenario 2: $r_1>r_2$}
The Haar gates in the \bw structure do not exhibit two-stage thermalization, but the dual-unitary $(1,1,a_z)$ gates can. The new ingredient is a “magnon” mode. While the local update rules for Haar random circuits in Eq.~\eqref{eq:K_move} can only move domain walls, the ones for dual-unitary circuits in Eq.~\eqref{eq:RDUCmove} permit swap processes $\ket{+-} \rightarrow b_{-} \ket{-+}$. This can create new pairs of domain walls, and, when bound, such a pair forms a magnon. Before $t_{\rm sat}$, pair creation means that a magnon, once formed, can only coexist with a (dressed) domain wall. Since the free energy of a magnon on top of a domain wall is always higher than a single domain wall, the single domain wall mode dominates, and $r_1=1$ $(2)$ for OBC (PBC) \cite{foligno_growth_2023}. However, once the domain wall can exit through the boundary after $t_{\rm sat}$, a standalone magnon can exist and compete with the domain wall. We believe this transition from a domain wall to a magnon mode creates the second stage with a smaller rate $r_2 < r_1$.


To confirm that magnon gives the rate $r_2$, we isolate the magnon contribution and compute a magnon (sub)partition function,
\begin{equation}
Z_{\text{mag}}(x,t) = \langle \cdots  +^* -^* +^* \cdots  +^* | \mathcal{\widehat{M}}(t) | - + \cdots +\rangle. \label{eq:Zmag}
\end{equation}
The initial state $\ket{-++\cdots+}$ (top boundary) is a domain wall on the left boundary, which is creating a magnon. The final states are constructed from the dual basis $\{\ket{+^*}, \ket{-^*}\}$ (note $|+\rangle$ and $|-\rangle$ are not orthonormal), which satisfies $\langle i^* | j \rangle = \delta_{ij}$, $i,j \in \pm$. These dual states pin the magnon at spatial positions $x=1,2,\cdots,L$. The partition function \eqref{eq:Zmag} contains all trajectories that start with a magnon at the left spatial boundary and end with a magnon at $x$, with possible branching and broadening corrections of the magnon mode in between. We recursively solve for these corrections from the numerical data of $Z_{\rm mag}(x,t)$ and resum them to obtain the asymptotic decay rate $r_{\rm mag}$ defined by $\sum_{x} Z_{\rm mag}(x,t) \sim \exp( - r_{\rm mag} t \ln q )$. Fig.~\ref{fig:schematics_r2az}(b) displays the numerical results for $r_{\rm mag}$ in random $(1,1,a_z)$ circuits, confirming the predictions of Eq.~\eqref{eq:r2_az_analy}. For PBC, $r_1=2$ always exceeds $r_{\rm mag}$, meaning that the magnon dominates in the second stage. For OBC, however, $r_1$ surpasses $r_{\rm mag}$ only when $a_z \geq 1/3$. These findings explain the physical mechanism of $r_2$, and clarify how $r_2$ behaves under different boundary conditions.

\begin{figure}[h]
\centering
\includegraphics[width=.38 \columnwidth]{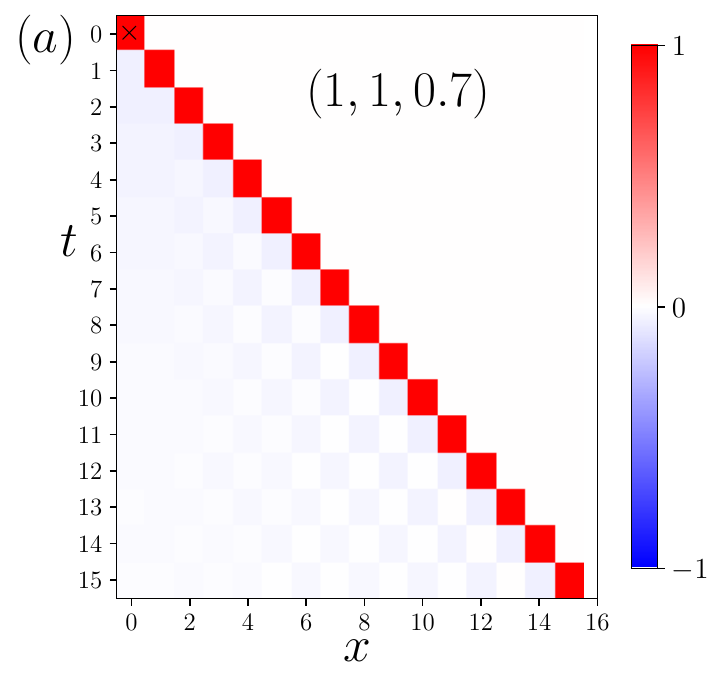} 
\includegraphics[width=.6 \columnwidth]{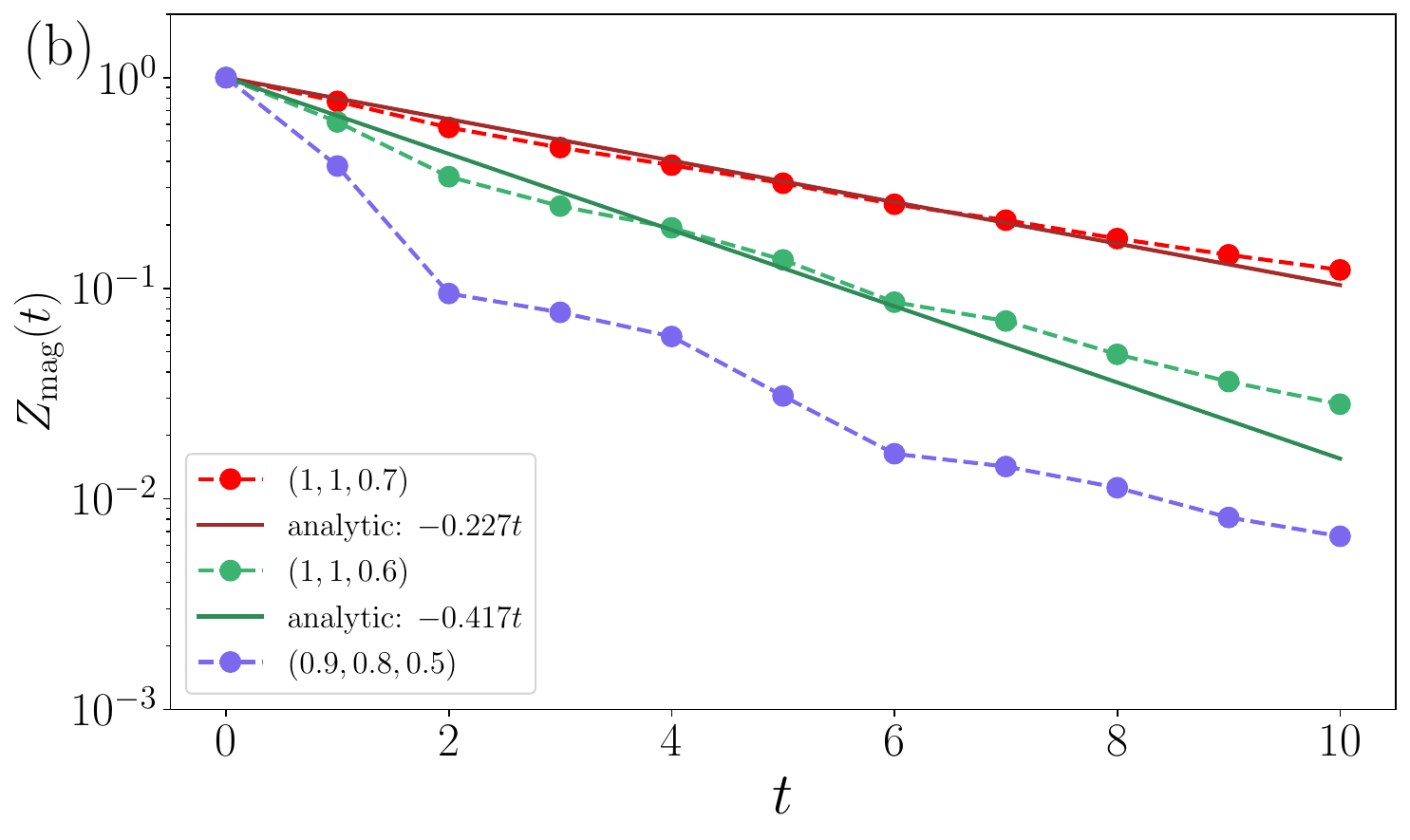} 
\caption{Magnon partition function and magnon decay rate. (a) heatmap of $Z_{\rm mag}(x, t) / \text{max}_x |Z_{\rm mag}(x, t)|$. The magnon mostly travels on the light cone. (b) Magnon decay rate for clean dual unitary circuits $(1,1,0.7)$, $(1,1,0.6)$ and clean non-dual unitary Floquet circuit $(0.9,0.8,0.5)$. }
\label{fig:Z_mag}
\end{figure}

\textit{Exact $r_{\rm mag}$ from correlation function.} We can exactly derive the analytic curve in Fig.~\ref{fig:schematics_r2az}(b) via dual unitarity, which confines local correlations to propagate strictly \textit{on} the light cone \cite{piroli_exact_2019}. From the heatmap of $Z_{\rm mag}(x,t)$ (Fig.~\ref{fig:Z_mag}(a)), the value at $x = t$ dominates for each $t$ and magnon is thus also mainly propagate on the $x=t$ ray. 
Instead of relaying on $Z_{\rm t, t}$, we introduce a modified partition function
\begin{equation}
\tilde{Z}_{\rm mag}(x, t ) = \langle + \cdots + - + \cdots + | \mathcal{\widehat{M}}(t) | - + \cdots +  \rangle. \label{eq:Zmodified}
\end{equation}
and seek for $\tilde{Z}_{\rm mag}(t,t)$. Contrary to $Z_{\rm mag}(x,t)$ in Eq.\eqref{eq:Zmag}, where the magnon was strictly pinned to position $x$, the partition function in Eq.~\eqref{eq:Zmodified} does not eliminate entirely but penalize other configurations with exponential cost. Thus it allows small $\mathcal{O}(1)$ fluctuations around a magnon at position $x$ in the final state and importantly does not affect the asymptotic decay rate $r_{\rm mag}$. Moreover, dual-unitarity reduces $\tilde{Z}_{\rm mag}(t,t) $ to an repeated applications of a quantum channel acting on the $\ket{-}$ state, as discussed in  \appref{app:AveragedChannel} and Ref.~\cite{piroli_exact_2019}. After averaging over single site, quantum channel is simplified and has only two eigenvalues: $\lambda_{+}=1$, which gives the dissipationless propagation of the $\ket{+}$ state, and $\lambda_-=2^{-r_{\rm mag}}$, where $r_{\rm mag}$ is the analytic expression in Eq.~\eqref{eq:r2_az_analy}. 


\textit{Generic systems} Importantly, this theory can be applied beyond random averaging to generic chaotic systems with either or both space and time translation invariance. As shown in Ref.~\cite{zhou_entanglement_2020}, going away from random averaging leads to bubble corrections that widen the domain walls in Fig.~\ref{fig:dw_S}(a) to have $O(1)$ width. The line tension function still after the renormalization by the bubbles\cite{zhou_entanglement_2020}. Similar corrections can occur for the magnon (\appref{sec:WZ}), so long as the bound state has lower energy than two separate domain walls. Otherwise, the magnon can dissolve through a bonding transition, as discussed in Ref.~\cite{nahum_real-time_2022-1}.


In Fig.~\ref{fig:Z_mag}, we numerically resummed the corrections to the magnon in a Floquet dual unitary circuit. It has parameter $(a_x, a_y, a_z) = (1,1,0.5)$ and a fixed single site unitary $u_1 = u_2 = u_3 = u_4 = e^{i(\sin(\phi)\sigma^x+\cos(\phi)\sigma^z)}$ with $\phi=0.6$. As in the average case, dual unitarity allows us to analytically solve $r_{\rm mag}$ from the eigenvalues of the quantum channel. 
The resummed value of $r_{\rm mag}$ even in systems of $L=12$ sites converge to the analytic prediction from channel calculation above, as shown in the inset of Fig.~\ref{fig:Z_mag}.

{\it Discussion} Our physical theory of the emergent domain wall and magnon has quantitatively explained all the observed finite system phenomenologies of $r_1$ and $r_2$: $r_1 < r_2$ is a geometric effect, while $r_2 < r_1$ is a 
magnon mode wining over the domain wall. 
Numerical evidence suggests that our theory works beyond random averaging and applies to generic (time-periodic) chaotic systems. 

Interestingly, there is a similar two-stage thermalization for a standalone magnon. Instead of starting from a domain wall, as in the case of the purity, we consider the square of a local correlation function, which after averaging over the operator creates a magnon state to begin with. If the magnon is in the center, $r_1$ will be the magnon decay rate as the domain wall cannot exist alone before $t_{\rm sat}$. For $t> t_{\rm sat}$, $r_2$ is once more determined by the competition between the domain wall and magnon. This is a \textit{reverse} transition from the magnon rates to whichever wins. 

If the domain wall decay rate is smaller, such a reverse transition can be used to measure the domain wall rate $\mathcal{E}(0)$ from $r_2$.
This is the case for $(1,1,a_z)$ circuits with $a_z<1/3$ after random averaging. An improvement of the protocol is to place the operator at the spatial boundary, so the magnon is instantaneously in a position to compete with the domain wall. In Fig.~\ref{fig:CtSqr}, we compute the boundary correlation function for dual unitary circuits $(1,1,a_z)$. The decay rate is $1$ when $a_z \lesssim \frac{1}{3}$, which is the domain wall rate. When $a_z \gtrsim \frac{1}{3}$ the decay rate is close to the ones obtained from the gap of the quantum channel, indicating it as the magnon rate. Our theory thus also works for the local correlation function. It provides a practical scheme to measure entanglement, a highly non-local quantity, from the decay of local observables. We anticipate its application in modern quantum simulation platforms and leave the practical implementation and study of noise to future works.

\textit{Acknowledgements}. C. J. thanks Tibor Rakovszky for his guidance on numerically implementing matrix product states. T. Z. was supported as a post-doctoral researcher from NTT Research Award No. AGMT
DTD 9.24.20. and the Massachusetts Institute of Technology. This work was supported by the US Department of Energy, Office of Science, Basic Energy Sciences, under Early Career Award Nos. DE-SC0021111 (C.J, under Prof. Vedika Khemani's grant). 
We acknowledge the accommodation of the KITP
program “Quantum Many-Body Dynamics and Noisy
Intermediate-Scale Quantum Systems” and the Simons Center for Geometry and Physics program "Fluctuations, Entanglements, and Chaos: Exact Results" in which parts of the work took place.



\let\oldaddcontentsline\addcontentsline
\renewcommand{\addcontentsline}[3]{}
\bibliographystyle{apsrev4-1}
\bibliography{update}
\let\addcontentsline\oldaddcontentsline





\appendix
\onecolumngrid





\section{REVIEW OF EXISTING RESULTS FROM TRANSFER MATRIX}
In this section, we tabulate numerical results of Refs.~\cite{bensa_fastest_2021,znidaric_solvable_2022} about the two-stage thermalization, specifically the decay rates of $Z_{L/2}(t) - Z( \infty)$ (they call it $|I_{L/2}(t) - I_{L/2}(\infty)|$). To ease comparison, we uniformize the notations to be consistent with this work. We also include results about the second stage decay in the in \stair geometry for dual unitary $(a_x, a_y, a_z)$, which is not covered in the main text. 

\begin{table}[h]
\centering
\begin{tabular}{ |c|c|c| } 
 \hline
     & $r_1$ & $r_2$ \\\hline
  \bw & $q^{ - \mathcal{E}(0) }$ &  smaller \\\hline
    \stair  & smaller  & $q^{ - \mathcal{E}(0) }$  \\\hline
\end{tabular}
\caption{Simplified summary of the two-stage thermalization process found in Refs.~\cite{bensa_fastest_2021,znidaric_solvable_2022}. $q$ is the local Hilbert space dimension, and $\mathcal{E}(v)$ is the line tension function. The decay rate transitions from $r_{1}$ to $r_{2}$ at $t_{\rm sat}$.}
\label{tab:bz_summary}
\end{table}

\begin{table}
\centering
\begin{tabular}{|c|c|c|c|c|c|c|c|c|c|} 
\hline
\multicolumn{3}{|c|}{\multirow{2}{*}{Model}}                 & \multirow{2}{*}{Figure} & \multicolumn{2}{c|}{$t_{\rm sat}$}                          & \multicolumn{2}{c|}{$r_1$}                                  & \multicolumn{2}{c|}{$r_2$}                                                            \\ 
\cline{5-10}
\multicolumn{3}{|c|}{}                                       &                         & \multicolumn{1}{c|}{$(n_A)$} & \multicolumn{1}{c|}{$(L_A)$} & \multicolumn{1}{c|}{$(\ln 2)$} & \multicolumn{1}{c|}{$(1)$} & \multicolumn{1}{c|}{$(\ln 2)$}                          & \multicolumn{1}{c|}{$(1)$}  \\ 
\hline
\multirow{7}{*}{\bw} & \multirow{4}{*}{PBC} & $(1,1,0)$       & 7(a)                    & $1/4$                        & $1/2$                        & $4$                            & $2$                        & $ 2 \log_2 3$                                           & $\log_2 3$                  \\ 
\cline{3-10}
                    &                      & $(1,1,0.5)$     & 7(b)                    & $1/4$                        & $1/2$                        & $4 $                           & $2$                        & $\approx 1.17 $                                         & $\approx 0.59$              \\ 
\cline{3-10}
                    &                      & $(1,1,a_z)$     & 9(a)                    & $1/4$                        & $1/2$                        & $4$                            & $2$                        & $-\log_2 |\lambda_2|$                                   & $\rightarrow /(2 \ln 2)$    \\ 
\cline{3-10}
                    &                      & $(0.9,0.8,0.5)$ & 15(a)(b)                & $1/4$                        & $1/2$                        & $\approx 2.77$                 & $\approx 1.39$             & $ \approx 1.34 $                                        & $\approx 0.68$              \\ 
\cline{2-10}
                    & \multirow{3}{*}{OBC} & $(1,1,0)$       & 8(a)(b)                 & $1/2$                        & $1$                          & $2$                            & $1$                        & $2 $                                                    & $1$                         \\ 
\cline{3-10}
                    &                      & $(1,1,0.5)$     & 8(c)(d)                 & $1/2$                        & $1$                          & $2$                            & $1$                        & $\approx 1.2$                                           & $\approx 0.6$               \\ 
\cline{3-10}
                    &                      & $(1,1,0.5)$     & 9(b)                    & $1/2$                        & $1$                          & $2 $                           & $1$                        & $\makecell{a < a^*: 2 \\ a> a_c: - \log_2 |\lambda_2|}$ & $\rightarrow /(2 \ln 2)$    \\ 
\hline
\multirow{5}{*}{S}  & \multirow{2}{*}{PBC} & $(1,1,0)$       & 10(a)(b)                & $1/2$                        & $1$                          & $2$                            & $1$                        & $2 $                                                    & $1$                         \\ 
\cline{3-10}
                    &                      & $(1,1,0.5)$     & 10(c)(d)                & $0.6 $                       & $1.2$                        & $2$                            & $1$                        & $\approx 0.60$                                          & $\approx 0.3$               \\ 
\cline{2-10}
                    & \multirow{3}{*}{OBC} & $(1,1,0)$       & 13(a)(b)                & $1$                          & $1/2$                        & $1$                            & $1/2$                      & $1$                                                     & $1/2$                       \\ 
\cline{3-10}
                    &                      & $(1,1,0.5)$     & 13(c)(d)                & $1$                          & $1/2$                        & $1$                            & $1/2$                      & $\approx 0.63$                                          & $\approx 0.32$              \\ 
\cline{3-10}
                    &                      & RUC             & 16                      & $1$                          & $1$                          & $ \log_2 \frac{3}{2} $         & $\log_2\frac{3}{2} /2$     & $2 \log_2 (5/4)$                                        & $\log_2 (5/4) $             \\
\hline
\end{tabular}
\caption{Detailed summary of the two-stage thermalization process found in various figures in Refs.~\cite{bensa_fastest_2021}. We list data in two sets of units: Using the convention of Refs.~\cite{bensa_fastest_2021}, the \bw circuit $(1,1,0)$ with PBC has $t_{\rm sat} = \frac{n_A}{4}$, $r_1 = 4 \ln 2$, $r_2 = 2 \log_2 3 \cdot \ln (2) = 2 \ln 3$, in our convention it has $t_{\rm sat} = \frac{L_A}{2}$, $r_1 = 2$ and $r_2 = \log_2 3 \cdot \ln (2) = 2 \ln 3$. Numerically $a^* \approx 0.32$.}
\label{tab:BZ_results}
\end{table}

Tab.~\ref{tab:bz_summary} is a simplified summary of $r_1$ and $r_2$. Tab.~\ref{tab:BZ_results} is a more detailed summary of the two rates. Ref.~\cite{bensa_fastest_2021} have several different conventions from our work. We list both in columns of Tab.~\ref{tab:bz_summary}: 
\begin{samepage}
\begin{itemize}
    \item Ref.~\cite{bensa_fastest_2021} defines one time step to consist of two layers of circuit evolution. In this definition the transfer matrix is time translation invariant. Their time $t_{\rm BZ}$ is twice of our time $t$: $t_{\rm BZ} = 2t$. 
    \item Ref.~\cite{bensa_fastest_2021} defines the decay rate to be $\exp(- r_{\rm BZ} t_{\rm BZ})$. It is related to our rate by $r_{\rm BZ} = 2  r \ln q$, where the factor of 2 comes from the change of time unit. 
    \item Ref.~\cite{bensa_fastest_2021} refer to the purity as $I(t)$ while we use $Z(t)$; Ref.~\cite{bensa_fastest_2021} denotes the linear system size as $n$ while we use $L$. 
\end{itemize}
\end{samepage}

\begin{figure}[] 
\centering 
\includegraphics[width=0.5\columnwidth]{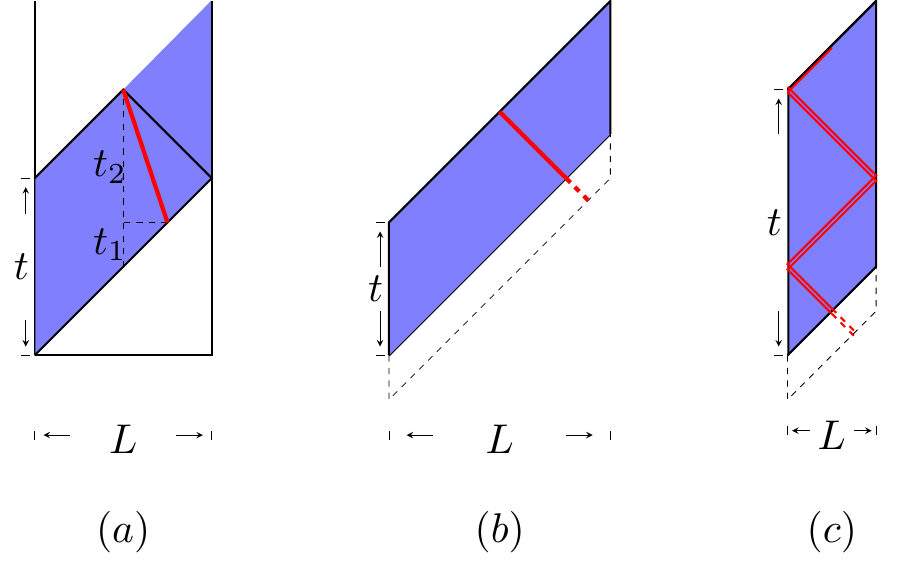} 
\caption{Domain wall and magnon in \stair geometry. (a) The time spans of each parts for $t < t_{\rm sat}$. If the (red) domain wall (anti)-slope is $v$, then $t_2 = \frac{t}{1+v}$. For dual unitary circuit, (b) domain wall favors to have $v = 1$, reducing its rate to be $\frac{\mathcal{E}(0)}{2}$; (c) magnon propagates on the light cone, also reducing its rate to $\frac{1}{2}$ of its value computed from the \bw geometry.}
\label{fig:fig_du_dw_magnon} 
\end{figure}

Our theory completely solves the results in Tab.~\ref{tab:bz_summary}. For \bw dual unitary circuits, $r_1 = 2$ for PBC and $1$ for OBC. $r_2$ also agrees with our prediction for the magnon mode $r_{\rm mag} = \ln \frac{3}{2-\cos \pi a_z}/\ln 2$. For \stair dual unitary circuit, $r_1$ is $r_1 = \frac{1}{2}$ for OBC and $1$ for PBC, exactly half of the value compared to the \bw geometry. This is again due to the geometry. For dual unitary circuits, the domain wall line tension if flat $\mathcal{E}(v) = 1$, so the minization $\min_v \frac{\mathcal{E}(v)}{1+v}$ selects out a domain wall on the light cone with $v = 1$, and $\min_v \frac{1)}{1+v} = \frac{1}{2}$, see Fig.~\ref{fig:fig_du_dw_magnon}(b). Further more both PBC and OBC has $r_2 \approx \frac{1}{2}r_{\rm mag}(a_z)$. Fig.~\ref{fig:fig_du_dw_magnon}(c) explains the origin of this $\frac{1}{2}$ factor: in dual unitary circuits, the magnon always travel on the light cone and it has slope $v = \pm 1$. This boundary effect is similar to Fig.~\ref{fig:fig_du_dw_magnon}(b). 
\section{MEMBRANE PREDICTION OF STAIRCASE GEOMETRY}

In the main text, we argue that the two decay rates are given by the minmial free energy of either a domain wall or a magnon. We assume the evolution is chaotic, and a membrane theory for the entanglement exists. 

The purity has one domain wall (membrane) that wanders from the top to a tilted bottom boundary, see Fig.~\ref{fig:dw_S}. To minimize the free energy, on one hand, the membrane tends to have larger $v$ to reduce the length; on the other hand, the membrane favors smaller $v$ for smaller line tension. The balance gives rise to the equilibrium value of $v^*$. The main difference between the regimes $t\leq t_{\rm sat}$ and $t> t_{\rm sat}$ is that in the former, the shortest path exits at the bottom (time) boundary, whereas after the saturation time, the shortest path exits at the spatial boundary. This gives rise to two different optimization problems. 

\begin{itemize}
    \item {\bf Case 1: $t\leq t_{\rm sat}$, $r_1$} 
\end{itemize}

The geometry of a tilted membrane is shown in Fig.~\ref{fig:dw_S}. Let the (anti)-slope of the membrane to be $v$. We have 
\begin{equation}
  t_1 + t_2 = t , \quad t_2 = v t_1,  \quad t_1  = \frac{t}{1+ v}
\end{equation}

Therefore the free energy is
\begin{equation}
F(v)\equiv \min_v \frac{\mathcal{E}(v)}{1 +v} \ln q  \label{eq:FreeEnergyDW}
\end{equation}

For random circuit, line tension function for the purity is given by \eqref{eq:ruc_tension}, hence the problem is that of minimizing \eqref{eq:FreeEnergyDW}. The result is velocity $v^* = \frac{(q-1)^2}{q^2+1}$, and the minimal free energy per unit length is
\begin{align}
F(v^*)&= \frac{1}{2} \ln \frac{q^2-q+1}{q}
\end{align}

Therefore $e^{-F(v^*) \ln q t} = \sqrt{\frac{q}{q^2 - q +1}}^t$  gives the ''phantom eigenvalue" $\lambda_{\rm ph}$ that relates to the decay rate as $r_1=-\ln|\lambda_{\rm ph}|$. This is consistent with the transfer matrix results from Ref.~\cite{znidaric_solvable_2022}.

\begin{itemize}
    \item {\bf Case 2: $t>t_{\rm sat}$, $r_2$}
\end{itemize}

When $t \gg t_{\rm sat}$, the entanglement has saturated, but $Z(t) - Z(\infty)$ still has exponentially small fluctuations. We therefore divide $Z(t)$ into two parts, one which approaches the static saturation value as $t \rightarrow \infty$, and one which is dynamical even past the saturation time,
\begin{align}
    Z(t)&=Z_{\lrbd}(t)+Z_{\bbd}(t). 
\end{align}

$Z_{\lrbd}(t)$ describes configurations that exit at the (left and right) spatial boundaries (Fig.~\ref{fig:dw_S}(c)),
and the difference $Z_{\lrbd}(t) - Z(\infty)$ is dominated by a domain wall ending at the bottom tip,
\begin{align}
    Z_{\lrbd}(t) - Z(\infty) \sim - e^{ -\mathcal{E}(t / (L/2) )t \ln q}.
\end{align}

$Z_{\bbd}(t)$, on the other hand, is dynamical. It describes configurations that exit at the bottom (Fig.~\ref{fig:dw_S}(d)), and is dominated by a vertical domain wall with $v=0$, 
\begin{align}
    Z_{\bbd}(t)\sim e^{-\mathcal{E}(0) t \ln q }.
\end{align}

For $t\gg t_{\rm sat}$, the effect of the tilted bottom boundary is increasingly negligible compared to the bulk contribution. Therefore, $Z(t) - Z(\infty)$ is dominated by $- e^{ -\mathcal{E}(t / (L/2) )t \ln q}+ e^{-\mathcal{E}(0) t \ln q } \sim  e^{-\mathcal{E}(0) t \ln q }$, which gives the decay rate $r_2 = \mathcal{E}(0)$. 



\section{MICROSCOPIC DESCRIPTION OF MARKOVIAN PROCESS OF AVERAGED PURITY}\label{app:microsopicDynamics}

The fundamental idea of this work and statistical mechanics in general is to make coarse-grained statements about physical observables based on microscopic rules. In the models considered here, the microscopic rules come from averaging over unitary gates according to the Haar measure and the resulting Weingarten calculus. We follow the main text notations and decompose a two-site gate as 
\begin{align}
u = (u_1 \otimes u_2 ) u_{\rm sym}(a_x,a_y,a_z)  ( u_3 \otimes u_4). 
\end{align}

We consider two ways to take random average: 
\begin{itemize}
    \item[(i)] Directly average over 2-site unitary $u$.
    \item[(ii)] Average over single-site rotations, keeping the $u_{\rm sym}$ parameters constant.
\end{itemize}
(i) gives the familiar Haar random dynamics, which have been extensively studied. These were the first quantum circuit models mapped to statistical mechanics models in seminal works like Refs ~\cite{nahum_quantum_2017,Nahum_2018,zhou_nahum_emergent_stat_mech2018,zhou_nahum_entanglement_membrane2019}.
This mapping generates the classical spins in our effective stochastic models; to obtain the update rules, one need to integrate out some intermediate spins variables. We state the key results here and highlight the difference from (ii).\\

For (i), the local transition matrix for the components of the two-site basis $\ket{s_i,s_{i+1}} \in \{\ket{++},\ket{+-},\ket{-+},\ket{--}\}$ is
\begin{align} 
M&= \begin{pmatrix} 1&K&K&0\\ 0&0&0&0\\ 0&0&0&0\\ 0&K&K&1 \end{pmatrix}, \label{eq:MHaar} 
\end{align} 
Here, transitions are governed by a single parameter $q$, representing the dimension of the local Hilbert space. The only non-trivial dynamics is a propagating domain wall, $\ket{+-} \rightarrow K(\ket{++} + \ket{--})$. 
Note that these gates are \textit{not} symmetric in time. When deriving $M$, in particular when integrating out the intermediate spin variables, an explicit time axis has to be chosen. The representation of $M$ in \eqref{eq:MHaar} is for evolution that reads backwards in time (from top to bottom in Fig.~\ref{fig:dw_S}). \\

For (ii), the average over single sites with fixed $u_{\rm sym}$ results in the following transition matrix: 
\begin{align} 
M&=\begin{pmatrix} 1& h&h&0\\ 0& b_+ &b_- &0 \\ 0& b_-&b_+&0\\ 0&h&h&1 \end{pmatrix}. \label{eq:MZnidaric} \end{align}

We have set $q = 2$ here and the transition rates depend on the 2-body couplings $a_x,a_y,a_z$ as follows: $h=(3-v)/9$, $b_{\pm} = (3 \pm 6u + 5v)/36$ and $u = \cos ( \pi a_x) $$+ \cos ( \pi a_y ) + \cos ( \pi a_z) $, $v = \cos( \pi a_x) \cos( \pi a_y ) + \cos( \pi a_y) \cos( \pi a_z ) + \cos( \pi a_z) \cos( \pi a_x )$ \cite{bensa_fastest_2021}. In addition to propagating, a local domain wall $\ket{+-}$ can transition to $\ket{-+}$, enabling the emergence of new, non-domain wall modes.


Using either microscopic rule set (i) or (ii), dynamical partition functions over spin configurations can be computed. The observable determines the boundary conditions of the partition function. The problem reduces to summing over all "paths" between the top and bottom configurations consistent with the local update rules. 
Importantly, weights like $b_+$ can be negative (it is for most $a_z$ values in the one parameter family $(1,1,a_z)$). Negative weights can also appear in the weights coming from the Weingarten functions in \cite{zhou_nahum_emergent_stat_mech2018} when we consider higher order moments. However, at larger length scales, there negative weights are part of the corrections to an overall positive transition rate. In short: The macroscopic process is stochastic, even if the individual $M$ gates are not. 


\section{NUMERICAL COMPUTATION OF PARTITION FUNCTION $Z(t)$}\label{app:Numerics}
The averaged purity of a half-system partition is represented by a domain wall at position $x$, evolved by a global Markovian \textit{operator} 
$\mathcal{\widehat{M}}(t)$, and a free boundary at $t=0$,
\begin{align}
    Z_{x}(t)= \frac{1}{(q^2+q)^L} \sum_{s=0}^{2^L-1} \langle s| \widehat{\mathcal{M}}(t)|x\rangle. \label{eq:Ztfreeboundary}
\end{align}
This is exactly the partition function defined in \eqref{eq:Z} of the main text. The free boundary condition is encoded by generic spin configurations $\ket{s} = \ket{s_1,s_2,\cdots,s_L}$, where $s_i=\{+,-\}$, and the domain wall $\ket{x} = |\cdots + +|_x - - \cdots \rangle$. To introduce the numerical techniques for purity calculations, we define the coordinate basis $\vec{s}$. Two important examples are: (i) the initial product state, written in Dirac notation as $(|\ket{0}+\ket{1})^{\otimes L}$, which maps to the coordinate vector $\vec{s}_0 = (1,1,1,\cdots,1)_{1\times 2^L}$; and (ii) the domain wall state, in Dirac notation $|\cdots + +|_x - - \cdots \rangle$, which maps to the vector $\vec{s}_{\text{DW}(x)}=(0,0,\cdots,1,0,\cdots,0)_{1\times 2^L}$, with a single 1 entry at the position given by unraveling the bitstring $\ket{x}$ as an integer. For example, the domain wall state $|+--\rangle$ maps to integer $4$, which unravels to $(0,0,0,1,0,0,0,0)$. We introduce this canonical basis because it is the basis used for numerical computations. Namely, we encode all the domain wall subsystem purities at time $t$ in a vector 
\begin{align}
    \mathcal{Z}(t)&= \sum_{s=0}^{2^L-1} Z_s(t) \ket{s} \label{eq:Zvec}
\end{align}
The evolution is given by the global Markovian \textit{matrix} $\mathcal{M}$, which acts directly on the coordinate basis,
\begin{align}
    \mathcal{Z}(t)&= \mathcal{M}\mathcal{Z}(t-1), 
\end{align}
and where we define $\mathcal{M}=\mathcal{M}_e \mathcal{M}_o$, with $\mathcal{M}_e = M_{1,2} M_{3,4} \cdots M_{L-1,L}$, and $\mathcal{M}_o = M_{2,3} M_{4,5} \cdots M_{L-1,L}$ for open boundary and $\mathcal{M}_o = M_{2,3} M_{4,5} \cdots M_{L,1}$ for periodic boundary. $\mathcal{M}(t)$ is defined at even and odd time steps as 
\begin{align}
    \mathcal{M}(2t) &= \mathcal{M}^t\\ 
    \mathcal{M}(2t+1) &= \mathcal{M}_o \mathcal{M}^t .
\end{align}
In the coordinate notation, \eqref{eq:Ztfreeboundary} can be written as
\begin{align}
    Z_x(t)&=\vec{s}_0^T \mathcal{M}(t) \vec{s}_{\rm DW}
    \label{eq:ZtfreeboundaryCoordNotation}
\end{align}
We highlight the difference between $\widehat{\mathcal{M}}(t)$ and $\mathcal{M}(t)$: The former is an operator that evolves the basis state in the effective spin basis, while the latter is a matrix that acts directly on the vector of coefficients $(Z_0(t),Z_1(t),...,Z_{2^L-1}(t))$. While this evolution is classical and admits a MPS representation, storing the full information generically scales exponentially with the system size $L$. 
For special cases like (i), the dynamics naturally restrict to tracking $O(L)$ components. But for more general cases like (ii), the space spans all $2^L$ partitions, and we have to leverage the MPS algorithm to truncate to the relevant subspace. As a further note on numerical conventions, the all-up and all-down purity states exhibit a one-way street behavior, in the following sense: Once a configuration assumes the all ``+'' or all ``-'' state, it cannot evolve further. The coordinates $Z_0(t)$ and $Z_{2^L-1}(t)$ 
herefore amass weight, while the other components of \eqref{eq:Zvec} decay exponentially. Since we are trying to capture this exponential decay, we subtract the all ``+'' and all ``-''  state at each time step in our numerics.\\

\textbf{Special case: (i) \eqref{eq:MHaar}} From the microscopic rules (i), we can easily write down a recursion relation for the coefficients $Z_x(t)$:
\begin{align}
    Z_x(t)&= K(Z_{x-1}(t-1)+Z_{x+1}(t-1)), \label{eq:(i)recursionOneStep}
\end{align}
where the subscript $x$ is a domain wall at location $x$. In a single time step, the domain wall can move either left or right with rate $K$. The crucial part is that despite repeated application of this stochastic evolution, we remain in the domain wall sector. As a result, we only need to track $L+1$ purities, keeping the computational complexity polynomial in $L$. We can rewrite \eqref{eq:Ztfreeboundary} as 
\begin{align}
    Z_x(t)&= \frac{1}{(q^2+q)^L} \sum_{x'=0}^{L+1} \langle x'| \mathcal{\widehat{M}}(t) |x\rangle
    =\vec{s}_0^T \mathcal{M}_{\rm DW}(t) \vec{s}_{\rm DW(x)}, 
\end{align}
where the sum is now restricted to only domain wall configurations $x'$ at $t=0$, which reduces the matrix matrix $\mathcal{M}(t)$ to the $(L+1) \times(L+1)$ domain wall subspace, denoted as $\mathcal{M}_{DW}(t)$, and the coefficient vectors $\vec{s}_0 = (1,1,\cdots,1)_{1 \times (L+1)}$ and $\vec{s}_{\rm DW(x)}=(0,0,\cdots,1,\cdots 0)_{(L+1) \times 1}$.\\

\textbf{Special case: (ii) \eqref{eq:MZnidaric}} For the microscopic rules of case (ii), the domain wall at position $(x,t)$ can undergo three possible transformations: it may decay into a domain wall at $(x \pm 1, t - 1)$, it can remain stationary at $(x, t - 1)$, or it might transition into a mode outside the domain wall sector, denoted as $\perp$. Mathematically, this is captured in the recursion relation below, 
\begin{align}
    Z_x(t)&= K(Z_{x-1}(t-1)+Z_{x+1}(t-1)) + b_+ Z_{x}(t) + b_- Z_{\perp}(t-1). \label{eq:(ii)recursionOneStep}
\end{align}
Repeatedly applying this recursion populates an increasing number of purity partitions that reside outside the domain wall sector, and the algorithm scales $\exp(L)$.
We leverage the MPS algorithm here, applying truncation techniques to states with low singular values. The bond dimension $\chi$ depicted in Fig.~\ref{fig:schematics_r2az} originates from this truncation process.
\section{SUBPARTITION FUNCTIONS AND GENERATING FUNCTION RESUMMATION}\label{app:subpartitionFunctions}
In this work, we have introduced several partition functions to understand the dynamics, including \eqref{eq:Z}, \eqref{eq:Zmag}, and \eqref{eq:Zmodified}. The full partition function \eqref{eq:Z} captures the exact purity decay. However, for efficient numerics and analytical understanding, it is useful to consider restrictions of the paths summed over. This defines a \textit{subpartition} function, which will always be such that the top and bottom boundary condition are a fixed and identical mode. The reason is that we can then treat the deviations from this mode perturbatively. 

Both our numerical simulations and analytic arguments based on free energy calculations indicate that the first stage of dynamics in dual unitary circuits is dominated by a domain wall mode;
so we consider a subpartition function 
\begin{equation}
Z_{\text{DW}}(x,t) = \langle  +^* +^*|_x-^* -^* \cdots \cdots | \mathcal{\widehat{M}}(t) | \cdots + +|_0 - - \cdots \rangle, \label{eq:ZDW}
\end{equation}
where $|+^*\rangle$ and $|-^*\rangle$ are the dual basis states, which we elaborate on and define in \appref{sec:dualbasis}. Importantly, they are used to pin a mode at the boundary, which cannot be done using the usual $\{\ket{+},\ket{-}\}$ states since those are not orthonormal. This partition function sums over all modes that start and end with a domain wall that travels a distance $x$ in time $t$, and is the main object of study in \cite{zhou_nahum_entanglement_membrane2019}. As shown in \cite{zhou_nahum_entanglement_membrane2019}, this domain wall contribution dominates even in dynamics that are not strictly Haar random, and thus sets the timescale for early-time entanglement growth. An important distinction should be made between $Z_{DW}(x,t)$ and the full partition function $Z_x(t)$ in Eq.~\eqref{eq:Ztfreeboundary}. $Z_x(t)$ starts with a domain wall at $x$, but it has a \textit{free} boundary conditions, and hence includes all possible modes. $Z_{DW}(x,t)$, on the other hand, is restricted to only domain wall contributions. Focusing on $Z_{DW}(x,t)$ allows us to isolate the effect of the dominant domain wall mode on entanglement dynamics.

\begin{figure}[h]
\centering
\includegraphics[width=0.7\columnwidth]{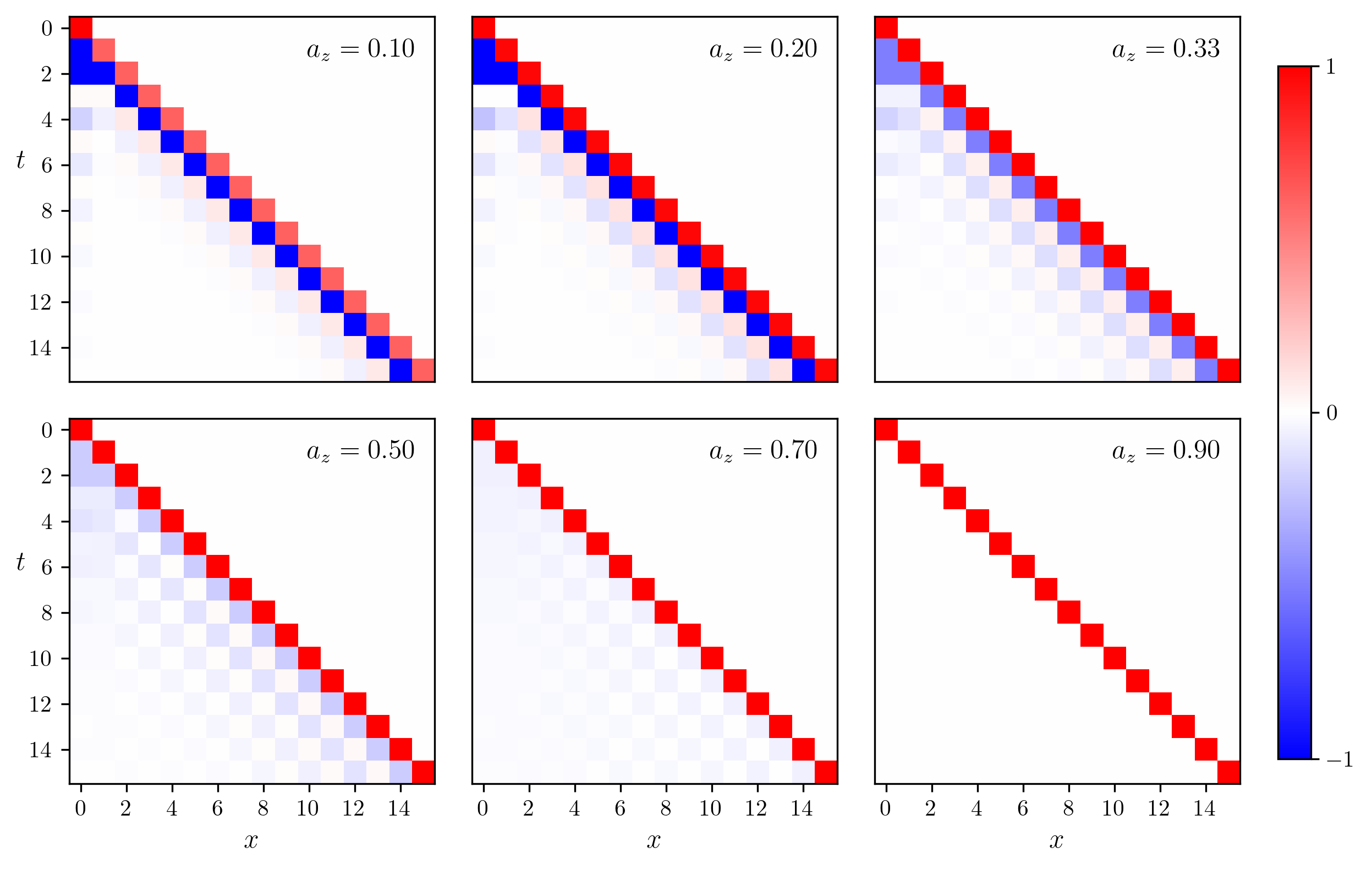}
\caption{$Z_{\rm mag}(x,t)$ for averaged dual unitary dynamics $(1,1,a_z)$, starting with a magnon mode $\ket{-+ \cdots +}$ from the top boundary. As $a_z$ increases, we observe the ballistic magnon mode becoming increasingly dominant in the dynamics, consistent with the symmetric unitarity $u_{sym}$ approaching a SWAP gate in the limit of $(1,1,1)$. For $(1,1,a_z \sim 0.2)$, the $b_-$ and $b_+$ moves have nearly equal magnitude but opposite signs, resulting in both a positive (red) and negative (blue) ballistic magnon.}
\label{fig:Zxt}
\end{figure}

After the domain wall decay, the dynamics transitions to a second stage, where a new slow mode dominates. Our numerics indicate that this stage is governed primarily by magnon modes propagating ballistically. Recall that the magnon move is determined by the coupling $b_-$ in Eq.~\eqref{eq:RDUCmove}, which acts as a swap on the $\ket{+-}$ state. As a first approximation, the decay rate is given by the ballistic magnon $Z \sim (b_-)^t$, which gives a decay rate $\ln(b_-)/\ln(2)$. However, comparison to numerical fits for the decay of $Z(t)-Z(\infty)$ show significant errors. To improve on this, we define an analog of \eqref{eq:ZDW} for the magnon mode, 
\begin{equation}
Z_{\text{mag}}(x,t) = \langle +^* -^*_x +^* \cdots \cdots +^*| \mathcal{\widehat{M}}(t) | - \cdots + -_0  +\cdots + \rangle.
\end{equation}
This magnon partition function includes the ballistic magnon $Z_{\rm mag}(t,t)=(b_-)^t$ at the bare level, but also allows for other (non-ballistic) magnon paths, and even intermediary trajectories that are not magnons at all. We show this magnon partition function in Fig.~\ref{fig:Zxt}. 
To carry out this sum, we decompose trajectories into irreducible diagrams $W(x,t)$, which start and end as a magnon, but are nowhere a magnon in between. The full partition function is then:
\begin{align}
Z(x,t) = W(x,t) + \sum_{t'<t} \sum_{y=1}^L Z(x,t') W(y-x,t-t'). \label{eq:WZrecursion}
\end{align} 
We have dropped the ``mag" subscript since this expansion is generic and exact for any perturbative expansion of a mode. However, it is only useful if two conditions hold: (i) the ratio $W(x,t)/Z(x,t)$ decays sufficiently fast, ideally exponentially in $t$, indicating that the irreducible weights decay rapidly, and (ii) the free-boundary partition function relates simply to $Z(x,t)$ with pinned boundary. Given $Z(x,t)$, we can solve for $W(x,t)$ iteratively, and verify $(i)$. This ensures that the decay rate $r_2$ we extract from $W(x,t)$ converges in time. 

\subsection{Computation of $r_2$ using Generating Functions}\label{sec:WZ}
\begin{figure}[h]
\centering
\includegraphics[width=0.7\columnwidth]{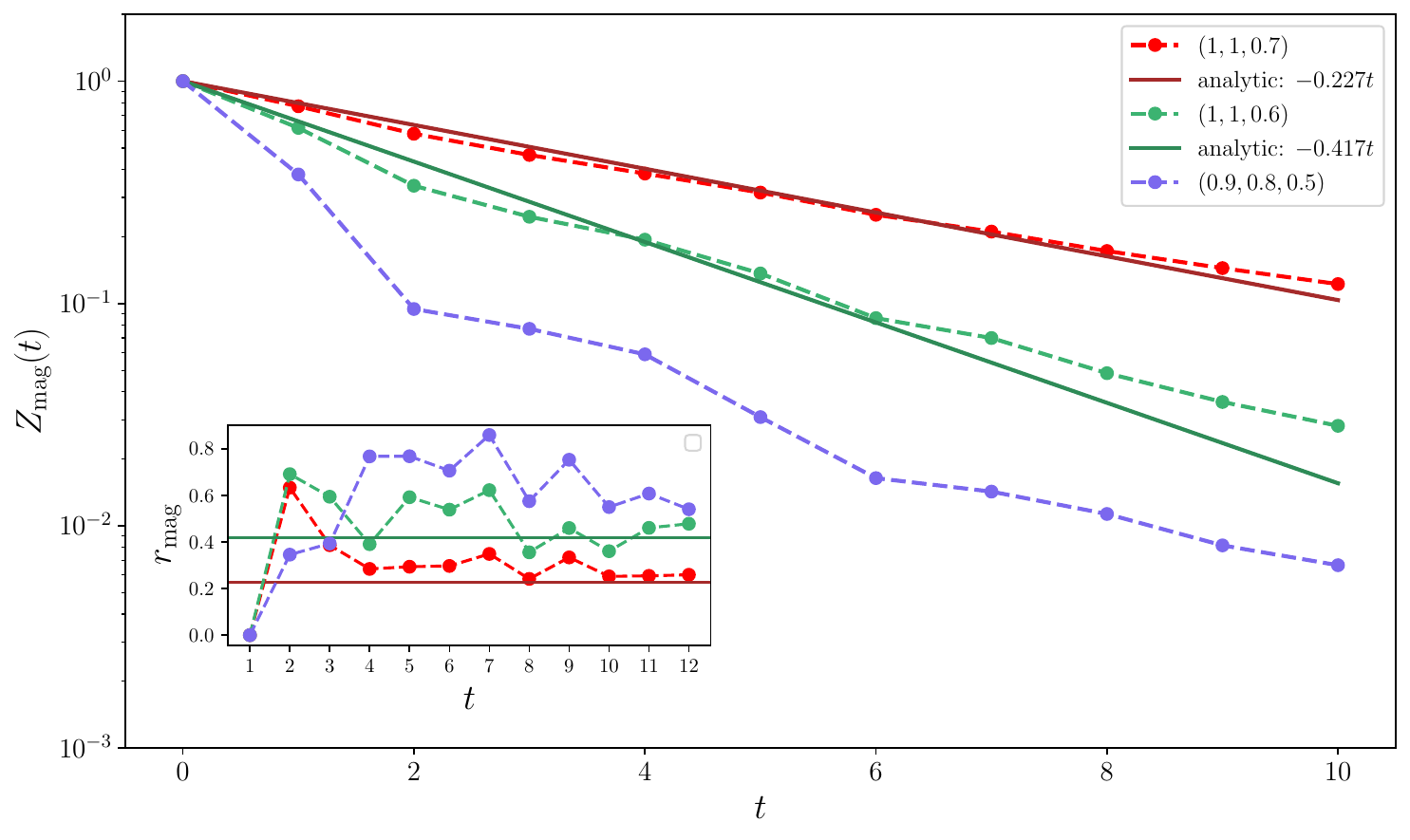}\\
\caption{Reproduction of Figure \ref{fig:Z_mag} data showing convergence of the $W/Z$ method, for two dual unitary circuits $(1,1,0.7)$ and $(1,1,0.6)$ and one non-dual unitary circuit $(0.9,0.8,0.5)$. The analytic prediction from the transfer matrix approach in Section \ref{app:AveragedChannel} is included for the dual unitary data (it does not apply to non-dual unitary circuits). The $W/Z$ method converges to the exact numerics very well, validating our method.}
\label{fig:ZxtwithWConvergence}
\end{figure}

We start with the recursion relation for $Z(t)$ as in \eqref{eq:WZrecursion}
\begin{align}
Z(t) &= W(t) + \sum_{t'<t} Z(t') W(t-t'), \label{eq:ZWt}
\end{align}
where we have dropped the position index $x$ for now. There are several ways to integrate out $x$ which we will discuss later, but our goal here is to derive a recursion in time $t$. To do this, we define the generating functions 
\begin{align}
\omega(x) &= \sum_{t=0}^{t_{\max}} W(t) x^t, \>\>
\varepsilon(x) = \sum_{t=0}^{t_{\max}} Z(t) x^t
\end{align}
We can thus rewrite \eqref{eq:ZWt} as $\omega(x) = \varepsilon(x) - \varepsilon(x)\omega(x)$,
ane solve for $\omega(x)$
\begin{align}
\omega(x) = \frac{\varepsilon(x)}{1+\varepsilon(x)}
\end{align}
Since we know $Z(t)$ decays exponentially as $Z(t) \sim e^{-at}$, the roots of $\omega(x)$ give the decay rate $a$. Specifically, if $x_0$ is a root of $\omega(x)$, then $a < \log(x_0)$, where the minimal root provides the tightest bound. We determine the roots numerically by finding $x_0$ such that $1-\sum_t W_t x_0^t =1 - W(1)x_0 - W(2)x_0^2 - \cdots - W(t)x_0^t = 0$. We verify the convergence of $r_2$ in Fig.~\ref{fig:ZxtwithWConvergence}.

\subsection{Integrating out $x$ in $W(x,t)$ and $Z(x,t)$}
\begin{figure}[h]
\centering
\includegraphics[width=0.7\columnwidth]{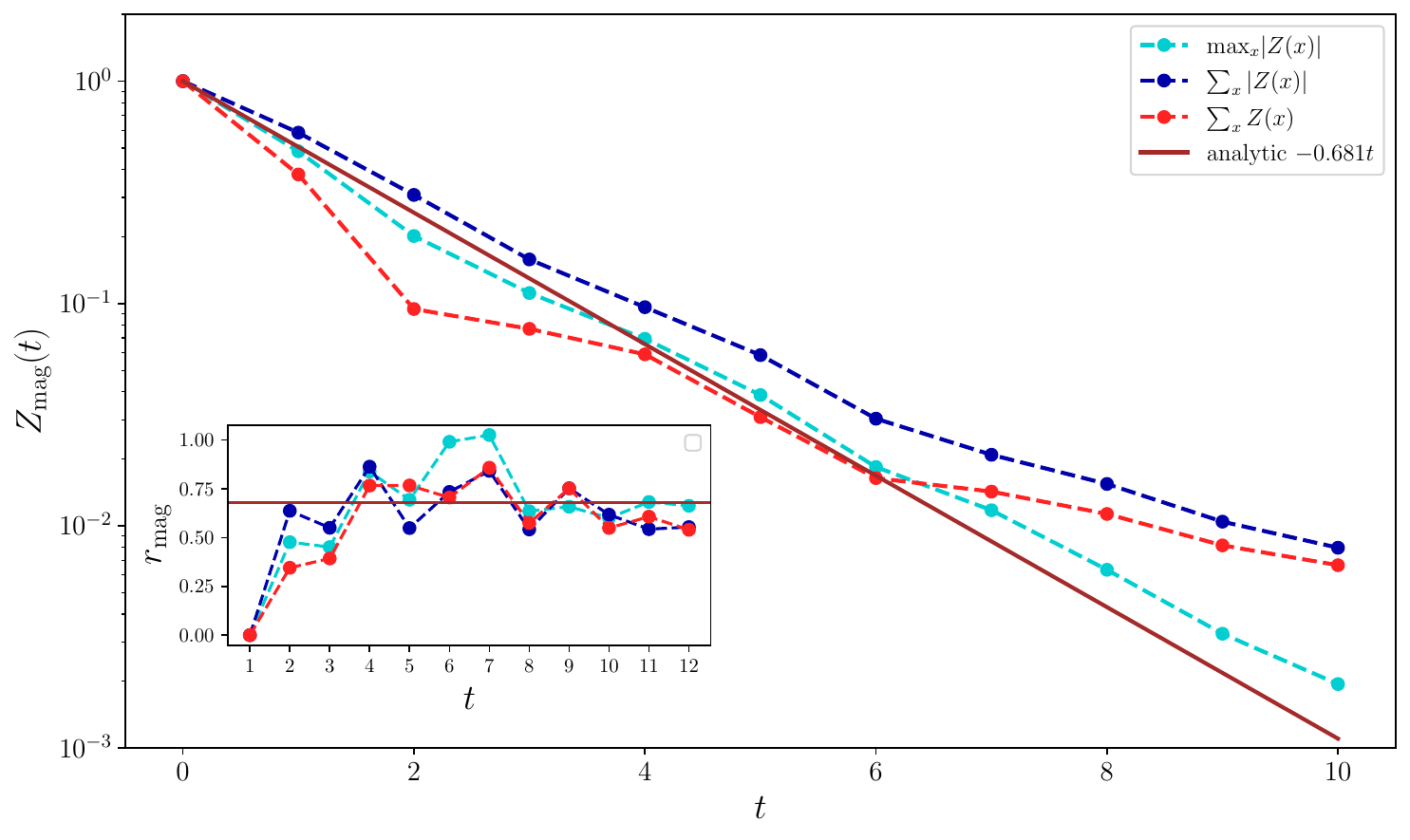}
\caption{Choices of integrating out spatial dependence $x$ in the partition function $Z(x,t)$, and convergence of $W/Z$. The main approaches are taking the maximum or sum over $x$ at each time $t$, with or without first taking the absolute value. While the maximum of $|Z(x,t)|$ shows fastest convergence, we use the proper choice of summing $Z(x,t)$ to extract $r_2$ in Fig.~\ref{fig:ZxtwithWConvergence} and Fig.~\ref{fig:schematics_r2az}(b) of the main text.}
\label{fig:ZxtIntegrateOutx}
\end{figure}
Here is a summary of the choices for integrating out the position dependence in $Z(x,t)$. There are two main approaches: 
\begin{itemize}
    \item \textbf{Maximum value over $x$ at each time $t$}
\end{itemize}
\begin{align}
Z(t) = \text{max}_x Z(x,t)
\end{align}
This tracks the envelope of $Z(x,t)$ over space.
\begin{itemize}
    \item \textbf{Sum over $x$ at each time $t$}
\end{itemize}
\begin{align}
Z(t) = \sum_x Z(x,t)
\end{align}
This averages over the spatial profile. An additional consideration is whether to take the absolute value before maximizing or summing over $x$, e.g. $Z(t) = \text{max}_x |Z(x,t)|$. The absolute value extracts the magnitude of the oscillations in $Z(x,t)$, which can be useful when the $b_+$ and $b_-$ moves in Eq. \eqref{eq:(ii)recursionOneStep} have equal magnitude but opposite sign, such as in the region with $(1,1,a_z \sim -0.2)$.

In Fig.\ref{fig:ZxtIntegrateOutx}, we show results for these four choices. While taking the maximum of the absolute value (Fig.\ref{fig:ZxtIntegrateOutx}, light blue curve) provides the best convergence to the analytic prediction of $r_2$, we adhere to the proper choice of summing $Z(x,t)$ as shown in Fig.~\ref{fig:ZxtwithWConvergence} and Fig.~\ref{fig:schematics_r2az} of the main text.

\subsection{Dual Basis}\label{sec:dualbasis}

Here we explain the dual basis used in \eqref{eq:Zmag}. The reason to introduce this basis is that the $\ket{+}$ and $\ket{-}$ states, which represent identity and SWAP contractions between the two copies, are \textit{not} orthonormalized. In fact, 
\begin{align}
    \langle + |+ \rangle = \langle - |- \rangle = q^2,\>\>  \langle + | - \rangle = \langle - | + \rangle = q
\end{align}
We thus define the dual basis through the condition $\langle \mu|\nu^*\rangle = \delta_{\mu, \nu}$. 
\begin{equation}
|+^*\rangle = \frac{1}{{q^2-1}} \left( |+ \rangle - \frac{1}{q} |- \rangle \right), \>\> |-^*\rangle = \frac{1}{{q^2-1}} \left( |- \rangle - \frac{1}{q} |+ \rangle \right) \label{eq:CanonicalDualChangeOfBasis}
\end{equation}

Note that the inner product now correctly satisfies $
\langle + | +^* \rangle = \frac{1}{{q^2-1}} \left( \langle + | + \rangle - \frac{1}{q} \langle + | - \rangle \right) = \frac{1}{{q^2-1}}(q^2-1) = 1$ and $\langle - | +^* \rangle = \frac{1}{{q^2-1}} \left( \langle - | + \rangle - \frac{1}{q} \langle - | - \rangle \right) = 0$, and similarly for $+ \leftrightarrow -$. This is the basis one uses to extract specific modes from the full partition function. 

\section{DUAL UNITARY QUANTUM CHANNEL}\label{app:AveragedChannel}
Dual-unitary circuits have the special property that local correlation functions propagate strictly on the light cone, as shown in previous works \cite{bertini_exact_2019-1}. This allows correlation functions to be calculated analytically through a transfer matrix formalism. To set up the following discussion, it is helpful to briefly review the behavior of two-point functions in dual-unitary circuits. As derived in \cite{bertini_exact_2019-1}, infinite-temperature two-point functions of local operators $a_0$ and $b_x$ in these circuits obey:
\begin{align} 
C^{ab}(x,t) & = \frac{1}{2^L} \Tr (a_0(t) b_x(0))  = \frac{1}{2^L} \sum_{\mu =l,r} \delta_{x,\pm t} \Tr(\mathcal{C}_{\mu}^{t} (a_0) b_x),
\end{align}
where $\mathcal{C}_{l/r}$ is a transfer matrix (quantum channel) that moves $a_0$ along either the left $\delta_{x,-t}$ or right $\delta_{x,t}$ light cone. It can be obtained directly from the dual-unitary two-qubit gates $U$ that make up the circuit. For concreteness' sake, we focus on the left-moving mode, but everything generalizes for $l \leftrightarrow r$. To be consistent with the convention of the main text, we take the time direction to be from top to bottom. The channel is is mathematically and pictorially defined as
\begin{equation}
\mathcal{C}_l(a)\equiv \Tr_1 (U^{\dagger} a \otimes I \>U) = \begin{tikzpicture}[baseline=(current bounding box.center)]
\dualgate[0][0]
\dualgate[0.2][0.07]
\node () at (-0.6,-0.6) {$\Tr_1$};
\node () at (-0.6,0.6) {$a$};
\draw (-0.6,0.7) -- (-.6,.8)  -- (-.5,.8) -- (-.28,.55);
\node () at (.8,0.6) {$I$};
\draw (-.49,-0.5) -- (-0.3,-0.43);
\draw (-.49+.99,-0.5+.99) -- (-0.3+.99,-0.43+.99);
\end{tikzpicture}.
\end{equation}
where $U$ is a dual unitary gate acting on qubits $1,2$, while $a$ is strictly local. In the folded picture, the identity operator $I$ and the trace over site 1 represent the same type of index contraction. To be concise, we use the folded notation going forward. As a cautionary note, we denote $I$ as the single-site Pauli matrix, while $\mathbb{1}$ represents the identity between two copies.

The correlations are then obtained analytically by diagonalizing the single-qubit quantum channel $\mathcal{C}_{l}$, which can be expressed in the Hermitian Pauli basis $\{I,X,Y,Z\}$. Since $\mathcal{C}_{l}$ is trace-preserving and unital (satisfying $\mathcal{C}_{l}(I) = I$), we only need to consider the traceless subspace spanned by $X$, $Y$ and $Z$. Denoting the three eigenvalues in this subspace by $\{\lambda_{\pm,i}:i=1,2,3\}$, for traceless operators $a_0,b_x$, we find 
\begin{align} C^{ab}_l(t) &=\sum_{i=1,2,3} c^{ab}_{l,i} \lambda_{l, i}^{t}\;, 
\end{align}
where the coefficients $c^{ab}_l$ are overlaps between the operators $a,b$ and the eigenmodes of $\mathcal{C}_{l}$.

The new insight is that the single $\ket{-}$ in a magnon state
is a good candidate for analysis using the transfer matrix formalism. It is a local mode, and, as numerics show, has the largest amplitude on the ballistic light cone. 
However, the magnon is intrinsically defined across two copies of the system. Therefore, we should compute the eigenmodes of the doubled channel $\mathcal{C}^2_l$. This is depicted below:

\begin{equation}
\mathcal{C}^2_l\ket{-}= \begin{tikzpicture}[baseline=(current bounding box.center)]
\newcommand{\distx}{1}
\newcommand{\disty}{0.5}

\dualgate[0][0]
\dualgate[0.2][0.07]
\dualgate[0+\distx][0+\disty]
\dualgate[0.2+\distx][0.07+\disty]

\draw (-.49,-0.5) -- (-0.3,-0.43);
\draw (-.49+.99,-0.5+.99) -- (-0.3+.99,-0.43+.99);

\draw (-.49+\distx,-0.5+\disty) -- (-0.3+\distx,-0.43+\disty);
\draw (-.49+.99+\distx,-0.5+.99+\disty) -- (-0.3+.99+\distx,-0.43+.99+\disty);

\draw (-0.5,0.5) .. controls (-0.5, 1) and (-0.6 + \distx, 1.05 + \disty) .. (-0.5 + 0.215 + \distx, 0.5 + 0.05 + \disty);

\draw (-0.31, 0.58) .. controls (-0.31, 0.88) and (-0.65 + \distx, 0.8 + \disty) .. (-0.5 + \distx, 0.5 + \disty);


\end{tikzpicture} \equiv
\begin{tikzpicture}[baseline=(current bounding box.center)]
\dualgatefat[0][0]
\node () at (-0.6,0.6) {$-$};
\node () at (0.6,0.6) {$+$};
\node () at (-0.6,-0.6) {$+$};
\end{tikzpicture}\label{eq:C2DisorderAverageMagnon}
\end{equation}
Note that $\mathcal{C}^2_l\neq \mathcal{C}_l\otimes \mathcal{C}_l$ when considering disorder averaged dynamics as in \ref{sec:ChannelDisorderAverage}. This is because the average couples the two copies, which happens \textit{before} we insert the $\{\ket{+},\ket{-}\}$ basis states that define the channel. 


\subsection{Disorder averaged systems}\label{sec:ChannelDisorderAverage}

In this scenario, we consider dynamics of type (ii) as specified in Sec.~\ref{app:microsopicDynamics}. The channel $\mathcal{C}_l^2$ only has one free input, which is the $\ket{+}$ or $\ket{-}$ state in the upper left corner of \eqref{eq:C2DisorderAverageMagnon}. For disorder averaged dynamics, the propagation of the two possible input states $\ket{++}$ and $\ket{+-}$ is given by the transfer matrix \eqref{eq:MHaar}. After propagating the state, we take the inner product with $\bra{+}$ on the left leg. In leading order approximation, we can use the ordinary basis $\ket{+},\ket{-}$ (rather than the dual basis) to extract the magnon state, which gives factors $\langle +|+\rangle= q^2$ and $\langle +|-\rangle=q$. We can summarize the action of this channel as
\begin{equation}
\begin{aligned}
\langle + | \mathcal{C}^2_l |++ \rangle / q^2  &=   | + \rangle \\
\langle + | \mathcal{C}^2_l |+- \rangle  / q^2 &= ( q b_{+}|+ \rangle  + b_{-} q^2 |- \rangle  + h q^2 |+ \rangle  + h q| - \rangle  ) /q ^2 .
\end{aligned}
\end{equation}
In matrix notation, 
\begin{equation}
\begin{pmatrix}
|+ \rangle \\
|- \rangle 
\end{pmatrix}
\rightarrow
\begin{pmatrix}
1 & 0 \\
\frac{b_+}{q} + h & \frac{h}{q} + b_{-}
\end{pmatrix}
\begin{pmatrix}
|+ \rangle \\
|- \rangle 
\end{pmatrix}
\end{equation}
One of the eigenvalue is $1$, which means it can propagate the $\ket{+}$ at no cost. The other is
\begin{equation}
\begin{aligned}
  \frac{h}{2} + b_{-}& = \frac{1}{12}(3- 2u + v ) = \frac{1}{12}( 3 - 2( -2 + \cos \pi a_z ) + (1 - 2 \cos \pi a_z ) )= \frac{2 - \cos \pi a_z}{3} 
\end{aligned}
\end{equation}
which in the correct units is exactly the decay of the magnon per unit step,
\begin{equation}
r_2=- \frac{\ln \frac{2 - \cos \pi a_z}{3}}{\ln 2}.
\end{equation}
\cite{bensa_two-step_2022} had conjectured this value, and we prove that it is the magnon decay rate.
\subsection{Non-disorder averaged systems}
\begin{figure}[h]
\centering
\includegraphics[width=0.7\columnwidth]{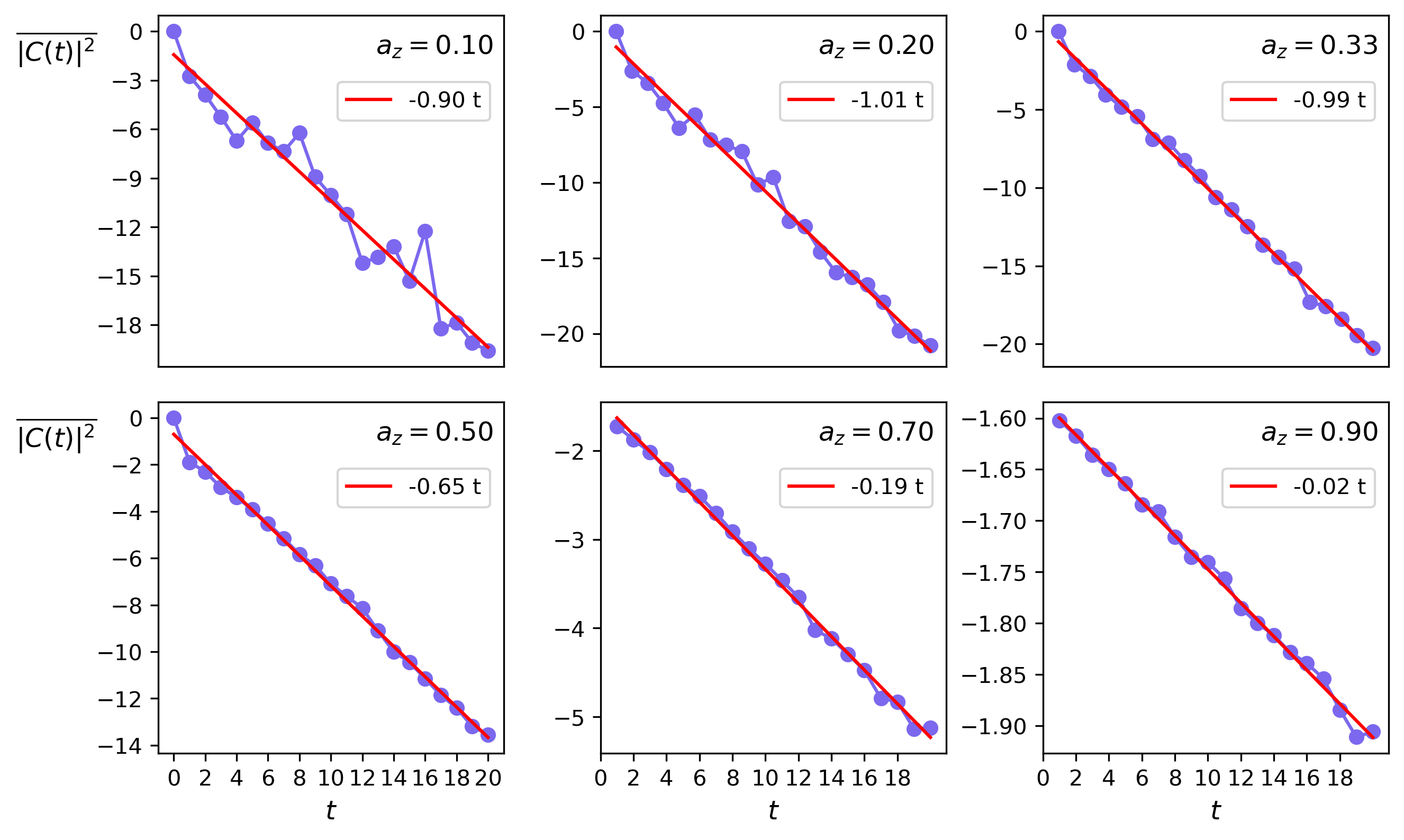}
\caption{Reverse transition. The correlation function $\overline{|C(t)|^2}=\overline{(X_1(t)|^2+|Y_1(t)|^2+|Z_1(t)|^2)/3}$, for dynamics in \eqref{eq:UFloquetZnidaric}, where the average $\overline{\cdots}$ is taken over initial product states. For $a_z <1/3$, we see (approximately) the domain wall decay rate.}
\label{fig:CtSqr}
\end{figure}
In our analysis thus far, we have made the key assumption that the magnon mode governs the decay rate of the partition function $Z_{\text{mag}}(t)$. However, considering the full partition function has open boundary conditions at the bottom, it is conceivable that the magnon could decay into other modes that actually determine the behavior. Furthermore, we have only examined the dynamics averaged over local $SU(2)$ rotations. To provide explicit validation that (i) the magnon does indeed dominate in the non-averaged case, and (ii) this behavior manifests even in single instantiations, we now transition to analyzing a fixed (non-averaged) circuit model. Without averaging, there are no effective pairing degrees of freedom $\{\ket{+},\ket{-}\}$, and the local update rules cannot be represented by the simple Markovian matrix equations derived earlier. 
To examine the non-averaged dynamics, we consider a translation invariant Floquet \bw model, specified by the local 2-site unitary
\begin{align}
     U&= e^{-i \pi/4 (XX +YY +a_z ZZ)}  u \otimes u, \>\>\>u=e^{i(\sin(\phi)\sigma^x+\cos(\phi)\sigma^z)}. \label{eq:UFloquetZnidaric}
\end{align}
This model was studied by Ref.~\cite{znidaric2023FloquetTwostep} with parameters $\phi=.6$ and $a_z=.5$, and we can directly compare our results. Note that we choose a model with translation invariance in space and time for two reasons: (1) spatial invariance ensues a position-independent magnon velocity, and (2) temporal invariance gives us access to the resummation methods in Sec.~\ref{sec:WZ}. 

Our approach is as follows. First, we numerically calculate the partition function $Z_{\rm mag}(x,t)$ using operator purities. The magnon state $|+^* +^* \cdots -^* +^* \cdots +^*\rangle$ in the dual basis maps to a linear combination of $2^L$ ordinary spin states, as given by Eq.~\eqref{eq:CanonicalDualChangeOfBasis}. Each overlap in the ordinary basis corresponds to an operator Renyi purity of the unitary evolution operator. Namely, the $\ket{+}$ sites lie in region $A$ while the $\ket{-}$ sites lie in region $\overline{A}$, the exact configuration being specified by the previously defined partition vector $\ket{s}$.
\begin{align}
    \langle \cdots +^* -^* +^* \cdots | U(t) \otimes U(t) |-+ \dots + \rangle &= \sum_s \langle s| U(t) \otimes U(t) |-+ \dots + \rangle =\sum_s \Tr_{A(s)} [\rho_{A(s)}(\ket{U(t)})]^2
\end{align}
where $\ket{U(t)}$ is mapping the evolution operator to a state.  The magnon overlaps reduce to a linear combination of $2^L$ operator Renyi purities. This algorithm is prohibitively costly as it scales exponentially in $L$. The limited system size also restricts the accessible time range for extracting the decay rate $r_2$. Since we want to avoid boundary effects from reflections, we constrain analyses to $t \leq L$ steps for a magnon initialized at $x=0$. As before, we apply the $W/Z$ technique of Sec.~\ref{sec:WZ} to determine $r_2$. The exponential decay of weight $W(t)$ enables predicting the asymptotic behavior despite the finite time window.


We have thus demonstrated that even without randomness, this perturbative calculation remains valid. The dual-unitary method allows reaching longer times to accurately extract the decay rate $r_2$.





\end{document}